\documentclass[10pt,titlepage]{article}
\usepackage[a4paper]{geometry}
\usepackage{textcomp}
\usepackage{amsmath}
\usepackage[onehalfspacing]{setspace}

\usepackage[a4paper]{geometry}
\usepackage{graphicx}

\usepackage{color,ulem}

\title{Non-specific binding of Na$^+$ and Mg$^{2+}$ to RNA determined by force spectroscopy methods}

\author{C. V. Bizarro$^{a,b,\dagger,**}$, A. Alemany$^{a,b,**}$, F. Ritort$^{a,b,*}$}

\date{ $^{a}$Departament de F\'{i}sica Fonamental, Universitat de Barcelona,
Diagonal 647, 08028 Barcelona, Spain; 
$^{b}$CIBER-BBN de Bioingenier\'{i}a, Biomateriales y Nanomedicina, 
Instituto de Sanidad Carlos III, Madrid, Spain; 
$\dagger$Present Address: Centro de Pesquisas em Biologia Molecular
e Funcional/PUCRS Avenida Ipiranga 6681, Tecnopuc, Partenon 90619-900,
Porto Alegre, RS, Brazil\\
\vspace{0.3cm}
*To whom correspondence should be addressed.
Tel: +34-934035869;
Fax: +34-934021149;
Email: fritort@gmail.com, ritort@ffn.ub.es\\
\vspace{0.3cm}
**These authors contributed equally to the paper
}

\begin{document}

\maketitle

\begin{abstract}
RNA duplex stability depends strongly on ionic conditions, and inside cells RNAs are exposed
to both monovalent and multivalent ions. Despite recent advances, we do
not have general methods to quantitatively account for the effects of monovalent
and multivalent ions on RNA stability, and the thermodynamic parameters for
secondary structure prediction have only been derived at 1M [Na$^+$]. Here, by mechanically
unfolding and folding a 20 bp RNA hairpin using optical tweezers, we study the RNA thermodynamics and kinetics at different monovalent and mixed monovalent/Mg$^{2+}$ salt conditions. We measure the unfolding and folding rupture forces and apply Kramers theory to extract accurate information about the hairpin free energy landscape under tension at a wide range of ionic conditions. We obtain non-specific corrections for the free energy of formation of the RNA hairpin and measure how the distance of the transition state to the folded state changes with force and ionic strength. We experimentally validate the Tightly Bound Ion model and obtain values for the persistence length of ssRNA. Finally, we test the approximate
rule by which the non-specific binding affinity of divalent cations at a given concentration
is equivalent to that of monovalent cations taken at 100 fold that concentration for small molecular constructs.
\end{abstract}

\section{Introduction}

RNA hairpins are elementary structures found in
many macromolecular assemblies. It is generally accepted that a
deeper understanding of their dynamics is a critical step towards
the elucidation of many biological processes, like the regulation of gene expression
\cite{Berkhout1989,Olsthoorn1999,Gollnick2005,Garst2009,Li2011}; the catalytic activity in many 
reactions \cite{Kawaji2008,Zhang2010}; the ligand-binding specificity \cite{Bardaro2009};
or the RNA folding problem \cite{Tinoco1999,Zhao2010}. DNA and RNA hairpins are
also appealing model systems for their simplicity as they are amenable
to exhaustive studies using a more physically-oriented approach, where
theoretical models can be rigorously tested using simulations and
experiments \cite{Chen2000,Hyeon2006}.

Many different and complementary biophysical methods
have been used to study these structures. For example, using time-resolved
nuclear magnetic resonance (NMR) spectroscopy and thermal denaturation experiments, kinetics and thermodynamics
of bistable RNA molecules were studied \cite{Petesheim1983,Freier1986,Walter1994,Serra1994,Serra1997,Xia1998,Mathews1999,Znosko2002,Mathews2004,Furtig2007}.
Recently, a photolabile caged RNA was designed to stabilize one ground-state conformation
and study the folding kinetics by NMR and CD spectroscopy under
different conditions, including Mg$^{2+}$ \cite{Furtig2010}. Laser
temperature-jump experiments have also been used to characterize the
folding kinetics of small RNA hairpins at the ns and $\mu$s timescales
\cite{Williams1989,Ma2006,Sarkar2009,Sarkar2010,Stancik2008,Zhang2006}. 
Using coarse-grained Go-like models, it was predicted that hairpins unfold in an
all-or-none process in mechanical experiments \cite{Hyeon2005}, in agreement with experimental
results \cite{Liphardt2001,Tinoco2004,Li2006}.

Within the cell, many dynamical processes
involving transient melting events of DNA and RNA double strands are
driven by the application of localized forces by molecular motors.
Therefore, single-molecule experiments are ideal to understand
the thermodynamics and kinetics of macromolecules inside cells \cite{Bustamante2004,Tinoco2010}.
As pointed out by Hyeon \textit{et al.} \cite{Hyeon2005}, force-denaturation using single-molecule experiments are intrinsically
different from thermally-induced denaturation: in bulk experiments
where the unfolded state is accessed by raising the temperature or
lowering the concentration of ions, the unfolded state is a high-entropy
state while in mechanical pulling experiments the unfolding process
is a transition from a low-entropy state to another
low-entropy state. Regions of the free energy
landscape normally inaccessible by conventional methods are probed
using mechanical experiments. Consequently, pathways
and rates of thermally-induced and mechanical unfolding processes
are expected to be different. 

In a previous work \cite{Manosas2006} we pulled an RNA hairpin using optical tweezers \cite{Huguet2010,Bustamante2006} 
to study the base-pairing thermodynamics, kinetics and mechanical properties at a fixed monovalent 
condition. A kinetic analysis was introduced to determine the location of the force-dependent kinetic
barrier, the attempt rate, and the free energy of formation
of the molecule. Here we performed a systematic study by mechanically pulling the same RNA hairpin
at different monovalent cation concentrations and also at mixed ionic
conditions containing different concentrations of Mg$^{2+}$ cations. 
This is important because
RNAs are highly charged polyanions whose stability strongly depends
on solvent ionic conditions \cite{Record1975,Record1976,Williams1996,Takach2004,Tan2005}.
Despite its biological significance, we have limited information
about RNA helix stability in mixed monovalent/multivalent ionic conditions
\cite{Tan2007,Tan2008}. In fact, the thermodynamic parameters for secondary
structural elements of RNAs have only been derived at the fixed standard
salt condition of 1M [Na$^{+}$] \cite{Walter1994,Serra1994,Serra1997,Xia1998,Mathews1999,Znosko2002,Mathews2004}.
Here we derived numbers such as
the persistence length describing the elastic
response of ssRNA and also the free energy of formation of an RNA hairpin at different monovalent and 
mixed monovalent/Mg$^{2+}$ conditions. Our results are compatible
with predictions obtained using the Tightly Bound Ion (TBI) model for mixed
ion solutions, that treats monovalent ions as ionic background and multivalent ions as responsible
from ion-ion correlation effects, and which takes into account only non-sequence-specific
electrostatic effects of ions on RNA \cite{Tan2005,Tan2007,Tan2008}. Our findings demonstrate the validity
of the approximate rule by which the non-specific binding affinity of divalent cations is equal to that
of monovalent cations taken around 100 fold concentration for small molecular constructs \cite{Heilman2001,TanChen2009}.

\section{Materials and Methods}

\subsection{Molecular Synthesis}

The RNA molecule was prepared as previously described \cite{Collin2005}.
Oligonucleotides CD4F (5'-AATTCACACG CGAGCCATAA TCTCATCTGG AAACAGATGAG
ATTA TGGCTCGC ACACA-3') and CD4R (5'-AGCTTGTGT GCGAGCCATA ATCTCATC TGTTTCCAGAT GAGATTATGGC TCGCGTGTG-3')
were anne\-a\-led and cloned into the pBR322 DNA plasmid
(GenBank J01749) digested with EcoRI (position 4360) and HindIII (position
30). The annealed oligonucleotides contain the sequence that codes
for a modified version of CD4-42F class I hairpin that targets the
mRNA of the CD4 receptor of the human immunodeficiency virus \cite{McManus2002}.
Oligonucleotides T7\_Forward (5'-TAATACGACTCA CTATAGG GACTGGTGA GTACTCA ACCAAGTC-3')
and T7\_Reverse (5'-TA GGAAGC AGCCCAGT AGTAGG-3') were used as primers
to amplify by PCR  a product of 1201 bp from the recombinant
clone containing the CD4 insert. This amplicon contains the T7 RNA
Polymerase promoter at one end, and was used as a template to synthesize
an RNA containing the RNA hairpin (20bp stem sequence and tetraloop
GAAA) and the RNA components of handles A (527 bp) and B (599 bp).
The DNA components of handles A and B were obtained by PCR from the
pBR322 vector (positions 3836-1 for handle A and positions 31-629
for handle B). Handle A was 3' biotinylated while handle B was tagged
with a 5' digoxigenin. Hybridization reactions were performed in
a formamide-based buffer \cite{Casey1977} with a step-cool temperature
program: denaturation at 85$^{\rm o}$C for 10 min, followed by
1.5 h incubation at 62$^{\rm o}$C, 1.5 h incubation at 52 $^{\rm o}$C,
and finished with a cooling to 10$^{\rm o}$C within 10 min.

\subsection{Measurement Protocol}

All experiments were performed using a dual-beam force measuring optical
trap \cite{Huguet2010,Bustamante2006} at 25$\pm$1 $^{\rm o}$C
in buffers containing 100 mM Tris.HCl (pH 8.1), 1mM EDTA, and NaCl
concentrations of 0, 100, 500, and 1000 mM, or in buffers containing
100 mM Tris.HCl (pH 8.1) and MgCl$_{2}$  concentrations of
0.01, 0.1, 0.5, 1, 4, and 10 mM. The monovalent cation concentration [Mon$^{+}$] includes 
the contributions from [Na$^{+}$] ions
and dissociated
[Tris$^{+}$] ions. At 25$^{\rm o}$C and pH
8.1, about half of the Tris molecules are protonated, therefore 100
mM Tris buffer adds 50 mM to the total monovalent ion concentration
\cite{Owczarzy2008}. Anti-digoxigenin polyclonal antibody-coated
polystyrene microspheres (AD beads) of 3.0-3.4 $\mu$m (Spherotech,
Libertyville, IL) were incubated at room temperature with the molecular
construct for 20 min. The second attachment was achieved inside the
microfluidics chamber using a single optically trapped AD bead previously
incubated with the RNA hairpin and a streptavidin-coated polystyrene
microsphere (SA bead) of 2.0-2.9 $\mu$m (G. Kisker GbR, Products
for Biotechnologie) positioned at the tip of a micropipette by suction
(Fig. 1A and 1B).

Tethered molecules were repeatedly pulled at two constant loading
rates of 1.8 pN/s or 12.5 pN/s by moving up and down the optical trap
along the vertical axis between fixed force limits and the resulting
force-distance curves (FDCs) were recorded (Fig. 2A). A pulling cycle
consists of an unfolding process and a folding process. In the unfolding
process, the tethered molecule is stretched from the minimum value
of force, typically in the range of 5-10 pN, where it is always at
its native folded state, up to the maximum value of force, typically
in the range of 25-30 pN, where the molecule is always unfolded. In
the folding process the molecule is released from the higher force
limit (unfolded state) up to the lower force limit (native folded
state) \cite{Liphardt2002}. A minimum of two molecules (different
bead pairs) were tested at each ionic condition, and a minimum of
100 cycles were recorded in each case (detailed statistics are given in the 
Supporting Material, section S1)

\subsection{Hairpin Model}

Under applied force it is feasible to reduce the configurational space
of an RNA hairpin containing $N$ base pairs (bps) to a minimum set
of $N+1$ partially unzipped RNA structures \cite{Manosas2006,Cocco2003,Manosas2005}.
Each configuration in this set contains $n$ adjacent opened bps in
the beginning of the fork followed by $N-n$ closed bps, with $0\leq n\leq N$.
The folded state (F) is defined as the configuration in which $n=0$
(all bps are formed), and the unfolded state (U) is the hairpin configuration
in which $n=N$ (all bps are dissociated). 
Based on a simple calculation (see Supporting Material, section S2) we conclude that fraying \cite{Woodside2006} plays a rather minor role (if any) on the folding/unfolding kinetics of the sequence under study (Fig. 1A) and we do not include it in our analysis. 
The stability of each configuration $n$
with respect to the F conformation is given by $\Delta G_{\rm n}(f)$, the free energy difference
at a given force $f$ between the duplex containing $N-n$ closed
bps and the completely closed configuration (F state),

\begin{equation}
\Delta G_{\rm n}(f)=\Delta G_{\rm n}(0)+\Delta G_{\rm n}^{ssRNA}(f)+\Delta G_{\rm n}^{d_{0}}(f).
\label{eq: free_energy}
\end{equation}

In eq. \ref{eq: free_energy} $\Delta G_{\rm n}(0)$ is the free
energy difference at zero force between a hairpin in the partially
unzipped configuration $n$ and a hairpin in the completely closed
configuration; $\Delta G_{\rm n}^{ssRNA}(f)$ is equal to the reversible
work needed to stretch the ssRNA strands of the hairpin in configuration
$n$ ($2n$ opened bases) from a random coiled state to a force-dependent
end-to-end distance $x_{\rm n}(f)$; and $\Delta G_{\rm n}^{d_{0}}(f)$
is the contribution related to hairpin stem orientation \cite{Mossa2009,Forns2011}.
An estimation of $\Delta G_{\rm n}(0)$ at 1M [$\mathrm{Mon^{+}}$]
can be obtained by using the nearest-neighbor (NN) energy parameters
widely employed to predict the stability of RNA secondary structures
\cite{Walter1994,Mathews1999}. It is given by the sum of the stacking
contributions of the duplex region, containing $N-n$ bps. The elastic
term $\Delta G_{\rm n}^{ssRNA}(f)$ is given by \cite{Mossa2009}

\begin{equation}
\Delta G_{\rm n}^{ssRNA}(f)=-\intop_{0}^{f}x_{\rm n}^{ssRNA}(f')df'.
\end{equation}

The molecular extension of ssRNA, $x_{\rm n}^{ssRNA}(f)$, can be
estimated using polymer theory (see section 2.4). Finally, the last
term in eq. \ref{eq: free_energy}, $\Delta G_{\rm n}^{d_{0}}(f)$,
is equal to the free energy of orientation of a monomer of length
$d_{0}$ along the force axis \cite{Forns2011}:

\begin{equation}
\Delta G_{\rm n}^{d_{0}}(f)=k_{\rm B}T\log\left[\frac{k_{\rm B}T}{fd_{0}}\sinh\left(\frac{fd_{0}}{k_{\rm B}T}\right)\right]\end{equation}

where $f$ is the applied force, $k_{\rm B}$ is the Boltzmann constant,
$T$ is the bath temperature, and $d_{0}$ is the diameter of a double
stranded chain, taken equal to 2 nm.

\subsection{Elastic models of ssRNA}

To model the elastic response of ssRNA we employed both the interpolation
formula for the inextensible Worm Like Chain (WLC) model and the Freely
Jointed Chain (FJC) model, which give the equilibrium end-to-end distance
$x$ of a polymer of contour length $l_{\rm n}$ stretched at a
given force $f$. These models have been mainly tested for long polymers. However, several studies indicate that they are generally applicable when the contour length is larger than the persistence length.
The inextensible WLC is given by:

\begin{equation}
f=\frac{k_{\rm B}T}{P}\left[\frac{1}{4\left(1-x/l_{\rm n}\right)^{2}}-\frac{1}{4}+\frac{x}{l_{\rm n}}\right]\end{equation}

where $k_{\rm B}$ is the Boltzmann constant, $T$ is the bath temperature
and $P$ is the persistence length \cite{Bustamante1994,Marko1995}.
The FJC model is given by

\begin{equation}
x=l_{\rm n}\left[\coth\left(\frac{fb}{k_{\rm B}T}\right)-\frac{k_{\rm B}T}{fb}\right]\end{equation}

where $b$ is the Kuhn length.

There are other models, such as the Thick Chain, that are more general than the WLC or the FJC and that have been used to fit the elastic response of biopolymers. Despite of their greater complexity, we do not expect a qualitative improvement of our results by using them.

\subsection{Kinetic Analysis}

We applied Kramers rate theory \cite{Kramers1940} to study the kinetics
of the transition between states F and U. The framework for understanding
the effect of an external force on rupture rates was first introduced
in \cite{Bell1978} and extended to the case where the loading force
increases with time \cite{EvansRitchie1997,Qian1999}. The assumption
that the transition state does not move under an applied force $f$
can be relieved by considering that the effective barrier that must
be crossed by a Brownian particle is force-dependent, $B_{\rm eff}(f)$.
The unfolding and folding rates can be obtained as the first passage
rates over the effective barrier,

\begin{subequations} 
\label{kinetic_rates}
\begin{equation}
k_{U}(f)=k_{0}\exp{\left(-\frac{B_{\rm eff}(f)}{k_{\rm B}T}\right)},
\label{eq: kinetic_rates_subeq1}
\end{equation}
\begin{equation}
k_{F}(f)=k_{0}\exp{\left(-\frac{B_{\rm eff}(f)-\Delta G_{\rm N}(f)}{k_{\rm B}T}\right)}.
\label{eq: kinetic_rates_subeq2}
\end{equation}
\end{subequations}

In eq. \ref{kinetic_rates}, F was selected as the reference
state and $\Delta G_{\rm N}(f)$ has been defined in eq. \ref{eq: free_energy}.
$k_{0}$ is the attempt rate for activated kinetics. The effective
barrier $B_{\rm eff}(f)$ can be obtained analytically from Kramers
rate theory (KT) \cite{Zwanzig2001,Hyeon2007} (detailed derivation provided in the Supporting
Material, section S3) as 

\begin{equation}
B_{\rm eff}^{KT}(f)=k_{\rm B}T\log\left[\sum_{n=0}^{N}h(n)\exp{\left(\frac{\Delta G_{{\rm {n}}}(f)}{k_{\rm B}T}\right)}\right]\label{eq: theor_effBarrier}\end{equation}

with $h(n)=\sum\limits _{n'=0}^{n}\exp{\left(-\frac{\Delta G_{\rm n'}(f)}{k_{\rm B}T}\right)}$.
Importantly, the location of the barrier along the reaction coordinate
can be obtained from the first derivatives of $B_{\rm eff}(f)$
with respect to force,

\begin{subequations} \label{eq: location_of_bef} \begin{equation}
x_{\rm eff}^{F}(f)=-\frac{dB_{\rm eff}(f)}{df},\label{eq: location_of_bef_subeq1}\end{equation}
 \begin{equation}
x_{\rm eff}^{U}(f)=\frac{d[B_{\rm eff}(f)-\Delta G_{\rm N}(f)]}{df}\label{eq: location_of_bef_subeq2}\end{equation}
 \end{subequations}

where $x_{\rm eff}^{F}(f)$ and $x_{\rm eff}^{U}(f)$ are the
distances from the effective barrier to the F and U states, respectively.
The force-dependent fragility parameter $\mu(f)$ \cite{Manosas2006},

\begin{equation}
\mu(f)=\frac{x_{\rm eff}^{F}(f)-x_{\rm eff}^{U}(f)}{x_{\rm eff}^{F}(f)+x_{\rm eff}^{U}(f)}\label{eq: fragility}\end{equation}

lies in the range [-1:1] and is a measure of the compliance of
a molecule under the effect of tension. Compliant structures deform
considerably before the transition event and are characterized by
positive values of $\mu(f)$, i.e. $x_{\rm eff}^{F}(f)>x_{\rm eff}^{U}(f)$.
In contrast, brittle structures are defined by negative values of
$\mu(f)$, $x_{\rm eff}^{F}(f)<x_{\rm eff}^{U}(f)$. A given
sequence can display different fragilities at different force regimes,
due to changes in the location of the transition state (TS) with force.

From the measured transition rates (see section 2.6) we can get estimators
for the effective barrier $B_{\rm eff}^{(U/F)}(f)$ for unfolding
and folding using the expressions in eq. \ref{kinetic_rates}:

\begin{subequations} \label{eq: barrier_method} \begin{equation}
\frac{B_{\rm eff}^{(U)}(f)}{k_{\rm B}T}=-\log k_{U}(f)+\log k_{0},\label{eq: barrier_method_subeq1}\end{equation}
 \begin{equation}
\frac{B_{\rm eff}^{(F)}(f)}{k_{\rm B}T}=-\log k_{F}(f)+\log k_{0}+\frac{\Delta G_{\rm N}(f)}{k_{\rm B}T}.\label{eq: barrier_method_subeq2}\end{equation}
 \end{subequations}

By comparing the experimental estimators of the kinetic barrier $B_{\rm eff}^{(U/F)}(f)$
with the effective barrier $B_{\rm eff}^{KT}(f)$ as predicted
by Kramers rate theory (eq. \ref{eq: theor_effBarrier}) we can extract
the free energy of formation of the hairpin $\Delta G_{\rm N}(0)$,
the attempt rate $k_{0}$ and the parameters that characterize the
elastic response of the ssRNA \cite{Manosas2006}. While $k_{0}$ always can be determined
by doing this comparison, there is a trade-off between the contributions
of the elastic response of the ssRNA and the free energy of formation
of the hairpin. Although this is not strictly true (the stretching
contribution term is force dependent whereas the free energy of formation
term is not) it holds to a very good degree. Therefore, if only the
free energy of formation of the hairpin is known \textit{a priori},
then we can extract the elastic properties of the ssRNA by matching
eqs. \ref{eq: barrier_method}a and \ref{eq: barrier_method}b with
eq. \ref{eq: theor_effBarrier}. On the contrary, if we only know
the elastic properties of the ssRNA, then we can extract the free
energy of formation of the hairpin (see Supporting Material, section S4).

\subsection{Data Analysis}

The molecular transitions during unfolding and folding can be identified
as force rips in a force-distance curve (FDC) \cite{Mossa2009}.
In order to extract the unfolding and folding rates (eq. \ref{eq: kinetic_rates_subeq1}
and \ref{eq: kinetic_rates_subeq2}) from experiments we have collected
the first rupture forces associated with the unfolding and folding
parts of each pulling cycle (Fig. 2A and 2B). By plotting the number
of trajectories in which the molecule remained at the initial configuration
(F state during the stretching part and U state in the releasing part
of the cycle) as a function of force $N(f_{i})$, divided by the total
number of trajectories $N_{0}$, we obtained experimental estimators
for survival probabilities $P_{U/F}(f_{i})=N(f_{i})/N_{0}$ of the
U and F states. Moreover, we obtained an experimental estimator for
the probability densities $\rho_{U/F}(f)$ of unfolding and folding
first rupture forces by doing normalized histograms of both datasets
($\rho_{U/F}(f)=\Delta N/(\Delta f\times N_{0})$, where $\Delta N$
is the number of events in the range between $f$ and $f+\Delta f$).
The survival probabilities are related to $\rho_{U(F)}(f)$ by the
following equations,

\begin{subequations} \label{eq: surv_probabilities} \begin{equation}
P_{U}(f)=1-\int_{f_{min}}^{f}\rho_{U}(f')\text{d}f',\end{equation}
 \begin{equation}
P_{F}(f)=1-\int_{f}^{f_{max}}\rho_{F}(f')\text{d}f'.\end{equation}
 \end{subequations}

If we assume a two-state transition, the time-evolution of the survival
probabilities is described by the following master equations \cite{Evans2009}:

\begin{subequations} \label{eq: master eqs.} \begin{equation}
\frac{\text{d}P_{U}(f(t))}{\text{d}t}=-k_{U}(f(t))P_{U}(f(t)),\end{equation}
 \begin{equation}
\frac{\text{d}P_{F}(f(t))}{\text{d}t}=-k_{F}(f(t))P_{F}(f(t)).\end{equation}
 \end{subequations}

With this assumption and the experimental estimators for survival
probabilities and densities, it is possible to extract the transition
rates $k_{U(F)}(f)$ from rupture force measurements using $k_{U(F)}(f)=-r\rho_{U(F)}(f)/P_{U(F)}(f)$,
with $r$ the pulling speed \cite{Manosas2006,Mossa2009,Evans2009}.

\subsection{Salt corrections}

It is interesting to experimentally measure the effect of salt on the free energy of formation of nucleic acid hairpins. However, UV absorbance experiments cannot be carried out for this particular sequence because its melting temperature is too high to obtain reliable results (see Supporting Material, section S5).
Therefore, as mentioned in section 2.3, the estimation of the free energy of formation of the RNA hairpin at 1M [Mon$^+$] is obtained using the NN energy parameters proposed by \cite{Walter1994,Mathews1999}. 
To introduce the effect of monovalent salt concentration [Mon$^{+}$] we assume a sequence-independent correction
$g_{1}([{\rm Mon}^{+}])$ for the free energy of formation of one
base pair. As
the free energy is measured relative to the F state we get for the
free energy correction of a hairpin with $n$ unzipped bps:

\begin{equation}
\Delta G_{\rm n}^{[\text{Mon}^{+}]}(0)=\Delta G_{\rm n}^{1M}(0)-ng_{1}([{\rm {Mon}^{+}])}\label{eq: monovalent_correction}\end{equation}

where $\Delta G_{\rm n}^{1M}(0)$ corresponds to the free energy
of formation of the $n^{th}$ configuration at 1000 mM [Mon$^{+}$]
at zero force. In the case of mixed monovalent/Mg$^{2+}$ conditions
we add a second sequence-independent correction term $g_{2}([{\rm {Mon}^{+}])}$
that captures the effect of Mg$^{2+}$ ions on the hairpin free energy
of formation:

\begin{equation}
\Delta G_{\rm n}^{[\text{Mon}^{+}],[\mathrm{Mg^{2+}}]}(0)=\Delta G_{\rm n}^{1M}(0)-ng_{1}([\text{Mon}^{+}])-ng_{2}([\mathrm{Mg^{2+}}]).\label{eq: mixed_cond_correction}\end{equation}

In what follows, unless stated otherwise, all monovalent and divalent 
salt concentrations $[\text{Mon}^{+}],[\mathrm{Mg^{2+}}]$ are expressed
in mM units.

\section{Results}

\subsection{Effect of force on thermodynamics and kinetics of an RNA hairpin}

We pulled the RNA hairpin at loading rates of 1.8 pN/s and 12.5 pN/s
in buffers containing different ionic conditions (see Materials and
Methods, section 2.2). From the unfolding and folding FDC we measured
the first rupture forces along many cycles. The resulting probability
distributions at each pulling speed and ionic condition can be found
in the Supporting Material (section S6). 
As previously observed \cite{Manosas2006}, this
hairpin displays a two-state behavior, with force jumps
signaling the transition between F and U states (Fig. 2A) and with no evidence of fraying or 
intermediate states (see Supporting Material, section S2). 

The order of magnitude of the resulting rupture forces and hysteresis effects are compatible with 
previous force-melting experiments carried out for other simple nucleic acid structures, like
P5ab and TAR RNA hairpins  \cite{Liphardt2001,Tinoco2004,Li2006} or short DNA hairpins \cite{Woodside2006}.
Moreover, results at 1 M [Na$^+$] were in significant agreement with solution predictions \cite{Walter1994,SantaLucia1998}.
As expected, hysteresis effects strongly depend on the loading rate, being 
lower at 1.8 pN/s and higher at 12.5 pN/s. This can be seen from the experimental 
distributions of unfolding and folding first rupture forces (Fig. 2B), which 
are closer for pulling cycles performed at 1.8 pN/s.

\subsection{Experiments at different monovalent cation concentrations}

We performed pulling experiments at four different NaCl concentrations
(see Materials and Methods, section 2.2). We find that 
the RNA duplex stability increases at higher [Mon$^{+}$] concentrations.
For instance, rupture force distributions are displaced
to higher forces (Fig. 3A) as we increase the concentration of NaCl.
The greater duplex stability at higher salt concentrations can also
be observed as an increase in the mean rupture force with the logarithm of the salt concentration (Fig. 3B). 
The standard deviation of rupture unfolding (folding) forces, that
are known to be proportional to $k_{B}T/x_{\rm eff}^{U(F)}(f)$
\cite{Mossa2009}, remain almost constant along the salt range explored.
That might denote that the position of the TS mediating the unfolding and the folding
transitions does not depend on [Mon$^{+}$], despite the fact that both transitions occur at higher forces. 
In Fig. 3C we see that the unfolding (folding) kinetic rates decrease (increase) with the salt concentration,
which again shows the stabilizing effect of salt on the RNA hairpin.

\subsection{Experiments at 1M NaCl}

From the current set of NN energy parameters for RNA secondary structures
obtained at 1000 mM NaCl concentration 
we can predict the free energy of formation $\Delta G_{\rm N}(0)$ for the RNA hairpin at this
particular condition using the mfold server \cite{Xia1998,Mathews2004,Tinoco1971,Zuker2003}.
We get $\Delta G_{\rm N}^{\rm Mfold}(0)=63.0$ $k_{\rm B}T$ (although our experiments were performed at 1050 mM of monovalent salt, we do not expect significant differences by comparing with the
prediction at 1000 mM).
By applying the kinetic method introduced
in section 2.5 we can evaluate the kinetic barrier associated to the unfolding reaction $B_{\rm eff}^{KT}(f)$ 
and $B_{\rm eff}^{(U/F)}(f)$
for a given elastic model for the ssRNA and find the one for which the theoretical
prediction by Kramers rate theory (eq. \ref{eq: theor_effBarrier})
best matches the experimental results (eq. \ref{eq: barrier_method}).
The procedure is shown in Fig. 4A and explained in the Supporting Material (section S4). We found the best fit to our data
using the inextensible WLC model with persistence length $P=0.75\pm0.05$
nm and interphosphate distance $a=0.665$ nm/base. The free energy
of formation obtained is equal to $\Delta G_{\rm N}(0)=65.3\pm0.3$
$k_{\rm B}T$, in reasonable agreement with the aforementioned value
for $\Delta G_{\rm N}^{\text{Mfold}}(0)$.

\subsection{Selection of elastic parameters: experiments with monovalent salts}

In order to know $\Delta G_{\rm N}(0)$ at monovalent ionic conditions
different from 1050 mM we need to know the effect of salt on the
elastic contribution of ssRNA strands. The elastic behavior of single-stranded
RNA (poly-U) has been studied in single-molecule stretching and fluorescence 
experiments carried out at various [Na$^{+}$] concentrations
\cite{Seol2004,Toan2006,ChenPollack2012}. 
Despite the extremely different contour length of the molecules under consideration 
(we are dealing with 44 bases-long chain in contrast to a polynucleotide of 1500-4000 bases in \cite{Toan2006}),
we take the values for the persistence length $P$
proposed in previous stretching experiments and add our value obtained at 1050 mM.
We assume that the elastic properties of ssRNA strands are independent of sequence, 
which can lead to a small error in the values obtained for the elastic properties of ssRNA strands
in the case of sequence-dependent behavior.
In fact, a 
sequence-dependent elastic behavior for ssDNA strands was previously considered as a possible explanation
for the specific salt corrections found for the NN energy values obtained from unzipping experiments
\cite{Huguet2010}. 
In the inset of Fig. 5 we plot the persistence lengths versus the monovalent salt concentration. To
fit the data, we employ the following dependence of $P$ on the
Debye screening length $\lambda_{D}^{\nu}$,

\begin{equation}
P\sim\lambda_{D}^{\nu}\propto\frac{1}{[\text{Mon}^{+}]^{\nu/2}}.
\label{eq: Persistence_length_Debye}
\end{equation}

No unique scaling law can be derived, as the value of $\nu$ can
be 1, 2 or $<$1, depending on the polymer properties \cite{Micka1996}.
We can fit all data using $\nu=1$ but also letting $\nu$ be
a free parameter, which results in a value of $\nu=0.6\pm0.06$. We
then interpolate both fits, in order to infer the values of $P$
in our experimental conditions of [Mon$^{+}$] concentration (50, 150, 550 and 1050 mM), as shown in Table 1.

\subsection{Monovalent salt correction to the free energies of formation of the
RNA hairpin}

Having obtained the elastic parameters that allow us to appropriately
describe the elastic response of ssRNA strands at different [Mon$^{+}$]
concentrations, we still need to characterize the effect of salt on
the energies of formation of the RNA hairpin at each intermediate
configuration $n$. It is generally assumed a sequence-independent correction
to the free energies of formation of nucleic acids duplexes \cite{SantaLucia1998,SantaLucia2004,Shkel2004}.
However, we have previously shown that a sequence-dependent salt correction
to the NN energy parameters of DNA improves the free energy prediction
of both unzipping and melting experiments \cite{Huguet2010}. Related
to this, it has been found that cation concentration affects RNA stability
in a sequence-dependent manner \cite{Vieregg2007}. In the absence
of RNA sequence-specific parameters available, we adopted a sequence-independent
salt correction (eq. \ref{eq: monovalent_correction}) given by $g_{1}([\text{Mon}^{+}])=m\log([\text{Mon}^{+}]/1000)$,
where [Mon$^{+}$] is expressed in mM units. As we will see, there are experimental and theoretical evidences that support the logarithmic effect of monovalent ions to the stability of nucleic acid hairpins.

Using this correction, the variation of $\Delta G_{\rm n}(0)$ with
monovalent salt concentration depends strictly on the value of the
constant $m$. In order to derive $m$ from our data, we compared
the estimators of $B_{\rm eff}(f)$ obtained experimentally ($B_{\rm eff}^{(U)}(f)$
and $B_{\rm eff}^{(F)}(f)$ in eq. \ref{eq: barrier_method}) with
the theoretical prediction ($B_{\rm eff}^{KT}(f)$ in eq. \ref{eq: theor_effBarrier})
at different values of $m$. In Fig. 4A-D, we see the correspondence
between theory and experiments at each monovalent ion concentration.
For all salt concentrations, we found the best agreement at $m=0.10\pm0.01$
kcal/mol. This value agrees with the sequence-independent salt correction reported
for DNA duplex oligomers in melting experiments, $m=0.110\pm0.033$
kcal/mol \cite{SantaLucia1998,Peyret2000}, and in unzipping experiments
of polymeric DNA, $m=0.104\pm0.010$ kcal/mol \cite{Huguet2010}.
Fig. 3D summarizes all the results. At a given force we see that the
height of the kinetic barrier increases with salt concentration, which
again indicates that salt increases kinetically the stability of the
RNA structure.

In Fig. 5 we show the dependence of the measured $\Delta G_{\rm N}(0)$
of the RNA hairpin on the monovalent ion concentration. 
As expected from earlier
observations on DNA \cite{Record1975,Dove1962,Privalov1969} and from
the application of counterion condensation theory to interpret polyelectrolyte
effects on equilibrium involving highly charged, locally rod-like polyelectrolytes
\cite{Record1976,Manning1972,Manning1978,Record1978}, we observe an approximately linear
dependence of RNA duplex stability on the logarithm of monovalent
salt concentration. 
Interestingly,
our data can also be well-described by the empirical expressions derived 
in \cite{Tan2005,Tan2007,Tan2008}, where the TBI model is used to predict the 
hairpin free energies at different ionic conditions (see Supporting Material, section S7).

By deriving the effective barrier as a function of force we can measure
the distance of the TS to the F and U states, $x_{\rm eff}^{F}(f)$
and $x_{\rm eff}^{U}(f)$ (eqs. \ref{eq: location_of_bef_subeq1}
and \ref{eq: location_of_bef_subeq2}), and the fragility $\mu(f)$
of the molecule as a function of the applied force (eq. \ref{eq: fragility}).
Fig. 6 shows the two extreme cases with 50 and 1050 mM [Mon$^{+}$]
(continuous and dashed lines respectively). In panel A we observe
that the location of the TS changes as a function of force. The same
trend is observed for the fragility in panel B, where the experimentally
measured points, the predicted force-dependent fragility (black curves),
and the expected values of the fragility for all possible locations
$n$ of the TS along the stem of the hairpin are represented (horizontal
grid, right scale). At low forces, the TS is located near the loop,
$n=19\pm1$ (Fig. 6A, dark gray curves). At intermediate forces that depend on salt concentration
the TS moves to the stem region $n=6\pm1$ (Fig. 6A, gray curves).
At large forces the TS has disappeared. These results are in agreement
with previous findings using the same hairpin sequence \cite{Manosas2006}.
Moreover, we see that at higher monovalent salt concentrations the
locations of the different TS mediating unfolding and refolding are
the same ($n\simeq$ 18, 6 or 0 for low, intermediate and high forces
respectively) but shifted to larger forces. 
These results agree with the Hammond's postulate \cite{Hammond1955}: at increasing
[Mon$^{+}$] the F state is increasingly stabilized while the TS is shifted towards
the U state; simultaneously, as force increases the TS approaches the F state.

\subsection{Experiments in mixed monovalent/Mg$^{2+}$conditions}

We have also performed pulling experiments in mixed monovalent/Mg$^{2+}$
buffers, containing a fixed concentration of Tris$^{+}$ ions (50
mM) and varying concentrations of Mg$^{2+}$(see Materials and Methods,
section 2.2). The rupture force distributions for all mixed monovalent/Mg$^{2+}$ conditions
can be found as Supporting Material (section S6). We found two regimes in the behavior
of the average rupture forces for unfolding and folding processes
along the range of [Mg$^{2+}$] experimentally explored. Below
0.1 mM [Mg$^{2+}$], there is no significant difference between
control (no Mg$^{2+}$ added) and magnesium-containing conditions
(Figs. 7A and 7B). However, at higher magnesium concentrations, we
found a linear dependence of average rupture forces with the logarithm
of [Mg$^{2+}$] (Fig. 7A and 7B). Interestingly, Owczarzy \textit{et
al.} have made a similar observation in DNA melting experiments done
in mixed monovalent/Mg$^{2+}$ conditions \cite{Owczarzy2008}. They
found that the ratio $R=\sqrt{[\text{Mg}^{2+}]}/[\text{Mon}^{+}]$
(both salts in molar units) is a convenient parameter to determine
whether divalent or monovalent ions are dominant on duplex stability.
If $R$ is less than 0.22, then monovalent ions are dominant and the
presence of Mg$^{2+}$ can be ignored. In our experiments, $R=0.0632$
and $R=0.2$ for the 0.01 mM and 0.1 mM [Mg$^{2+}$] conditions,
respectively. As in the case of pure monovalent ion conditions, the
standard deviation of rupture forces remains almost constant and we
also observed a linear dependence of $\log{k_{U}(f)}$ and $\log{k_{F}(f)}$
on the applied force for the different [Mg$^{2+}$] tested (Fig.
7C).

In order to obtain the free energy of formation $\Delta G_{\rm N}(0)$
of the RNA hairpin at different magnesium concentrations, we employed
the empirical expression derived in \cite{Tan2007,Tan2008} by applying 
the TBI model to predict the RNA helix stability in mixed monovalent/Mg$^{2+}$ 
ionic conditions (see Fig. 9, inset). Using this mixed salt correction, it is possible
to obtain the sequence-independent correction of one base pair
$g_{2}([\text{Mg}^{2+}])$ at any mixed salt condition using eq. \ref{eq: mixed_cond_correction}:

\begin{equation}
g_{2}([\text{Mg}^{2+}])=\frac{1}{N}\left(\Delta G_{\rm N}^{{\rm Mfold}}(0)-\Delta G_{\rm N}^{{\rm TBI}}(0)-Nm\log([\text{Mon}^{+}/1000])\right).
\label{eq: g2}
\end{equation}

From this expression, we can extract the value of $\Delta G_{\rm n}^{[\text{Mon}^{+}],[\mathrm{Mg^{2+}}]}(0)$
in eq. \ref{eq: mixed_cond_correction} for any intermediate state
$n$. By varying the mixed salt-dependent values for the persistence
length $P$ of the ssRNA for each [Mg$^{2+}$], we can now determine
the value of $P$ that results in better agreement between the predicted
effective barrier $B_{\rm eff}^{KT}(f)$ and our experimental estimations
(Fig. 8 and Supporting Material, section S4). All the results are summarized in Fig. 7D, where we can
see that the stability of the hairpin increases with magnesium concentration.
The dependence of $P$ on [Mg$^{2+}$] is shown in Fig. 9. Table
2 summarizes the results obtained for the persistence length $P$
and the attempt frequency $k_{0}$. The position
of the TS varied with [Mg$^{2+}$] in a way similar to what we
found for [Mon$^{+}$]. In a specified force the TS mediating
the unfolding and folding transitions is shifted toward the U
state as the [Mg$^{2+}$] is raised (Fig. 10A), in agreement
with the Hammond's postulate \cite{Hammond1955}.
The force-dependence of the position of the TS with respect to the
F state $x_{\rm eff}^{F}(f)$ and the hairpin fragility $\mu(f)$
are similar in both monovalent and mixed ionic conditions (Figs. 6A, B
and 10A, B). At low forces the TS is located near the loop, whereas
at intermediate forces it is located in the stem region $n=6\pm1$
(Fig. 10A, B).

\section{Discussion }

The effect of monovalent
ion concentration on DNA stability has been extensively studied and
there is a variety of empirical salt corrections available in the
literature \cite{Owczarzy2004,Jost2009}. There is no general agreement
about the accuracy and limitations of use of salt corrections in terms
of sequence length and range of salt concentrations \cite{Owczarzy2004}.
Recently, we have reported 10 NN salt correction parameters for prediction
of DNA duplex stability derived from single-molecule experiments \cite{Huguet2010}.
However, there is no equivalent study on RNA duplexes and the experimental
data available for the salt effects on RNA duplex stability are limited
to short sequences that display a two-state behavior \cite{Vieregg2007}.
Different polyelectrolyte theories try to characterize the interaction between counterions and nucleic acids to study ionic effects to the molecular stability. The most accepted are mean field theories such as the Poison-Boltzmann and the counterion condensation theories \cite{Manning1972,Manning1978}. Recently, the Tightly Bound Ion (TBI) model has been proposed \cite{Tan2005}. It incorporates correlation and fluctuation effects for bound ions, and has been extended
to treat RNA helices under mixed monovalent/divalent salt conditions \cite{Tan2007,Tan2008}. It was shown that the
TBI improves the prediction of the stabilities of RNA duplexes smaller than 15 bp \cite{Tan2007,TanChen2006}. 

Here, we performed a detailed characterization of the effect of monovalent
and mixed monovalent/magnesium concentrations on the stability of a
RNA hairpin containing a stem of 20 bp by mechanically unfolding
and folding the molecule using optical tweezers. The results we have
obtained can be very well described by the empirical formulas derived from
the TBI model for predicting the stabilities of RNA hairpins in monovalent
ion and mixed monovalent/Mg$^{2+}$ buffer conditions. A comparison of the counterion condensation theory \cite{Manning2002} is provided in the Supporting Material (section S8), and although an acceptable accuracy is found at monovalent salt conditions, we observe that correlation effects become important in the presence of Mg$^{2+}$ ions and therefore the TBI gives improved predictions. 

When a hairpin is melted mechanically the U state is subjected to
a tension. In standard optical melting and calorimetric experiments,
the U state is under zero force. Therefore, to appropriately compare
both phenomena, it is necessary to discount the energy required to
stretch the single stranded unfolded state in mechanical experiments.
However, we have limited information about the elastic properties
of single stranded RNA as compared to DNA. Despite the chemical similarity
between both polymers, the RNA-specific 2'-OH group favors a conformationally
restricted C$3'-endo$ structure \cite{Rich2003}. Moreover, in previous
single-molecule studies \cite{Seol2004,ChenPollack2012}, the elastic response
of a single-stranded poly-U RNA was better described by a WLC model
rather than by a FJC model. Based on these considerations, we decided
to use the numbers derived from single-molecule stretching experiments
for the elastic behavior of ssRNA molecules (as explained in section
3.2) to appropriately account for the energetic contribution of stretching
the ssRNA strands during mechanical unfolding.

By applying the Kramers rate theory and correctly accounting for the
elastic contribution of ssRNA stretching we were able to obtain the
free energy of formation of the RNA hairpin from pulling experiments
done in non-equilibrium conditions. This allowed us to obtain a large
number of trajectories at different ionic conditions in a feasible
timescale. In this way, we obtained the effective barrier of the unfolding
reaction as a function of force for different ionic conditions.

The ability of Mg$^{2+}$ ions to stabilize RNA structures at much
lower concentrations than monovalent ions was recognized almost forty
years ago \cite{Cole1972}. In fact, by plotting the values of the
free energies of formation versus salt concentration (expressed in
mM units) we can collapse data for both types of salt into a single
master curve by multiplying [Mg$^{2+}$] by a factor 100 (Fig.
11A). This effect has been previously observed \cite{Williams1989,Heilman2001,TanChen2009,Schroeder2000,Shankar2006} and can be explained using the counterion condensation theory which account 
for strong correlations between counterions and polyelectrolytes 
\cite{Heilman2001,Manning1978, Record1978, Shklovskii1999}. A similarity (but not a data
collapse) is obtained for the persistence length values of ssRNA (Fig. 11B). 
It can be interpreted as the screening effect of the counterions present
in solution: the longer the screening length, the stiffer the molecule.
This equivalence found between the non-specific
binding of monovalent and divalent salts with nucleic acids might
be useful to develop new biochemical assays in situations where divalent
ion-specific interactions could be detrimental \cite{Douglas2009}.

As a further step, it would be very interesting to extend this study
by mechanically unfolding an RNA hairpin containing a Mg$^{2+}$-specific
binding site using our experimental setup. Eventually, the energetic
contributions of different specific binding sites could be dissected
and incorporated into current models of RNA structure prediction.

\section{Supporting material}

Supporting material is available at NAR Online.

\section{Funding}

This work was supported by the Human Frontier Science Program (RGP55-2008);
the Instituci\'o Catalana de Recerca i Estudis Avan\c{c}ats;
the Centro de Investigaci\'on Biom\'edica en Red en Bioingenier\'ia, Biomateriales y Nanomedicina;
and the Spanish Ministerio de Educaci\'on-Ministerio de Ciencia e Innovaci\'on (AP2007-00995 to A. A., FIS2011-19342 to F. R.).

\section*{Acknowledgements}

We acknowledge I. Tinoco, M. Manosas and A. Bosco for a careful reading
of the manuscript; We also acknowledge R. Eritja and A. Avi\~n\'o for their advice and help in carrying out the UV absorbance measurements.

\clearpage{}

\section{Figure Legends}

\includegraphics[scale=1]{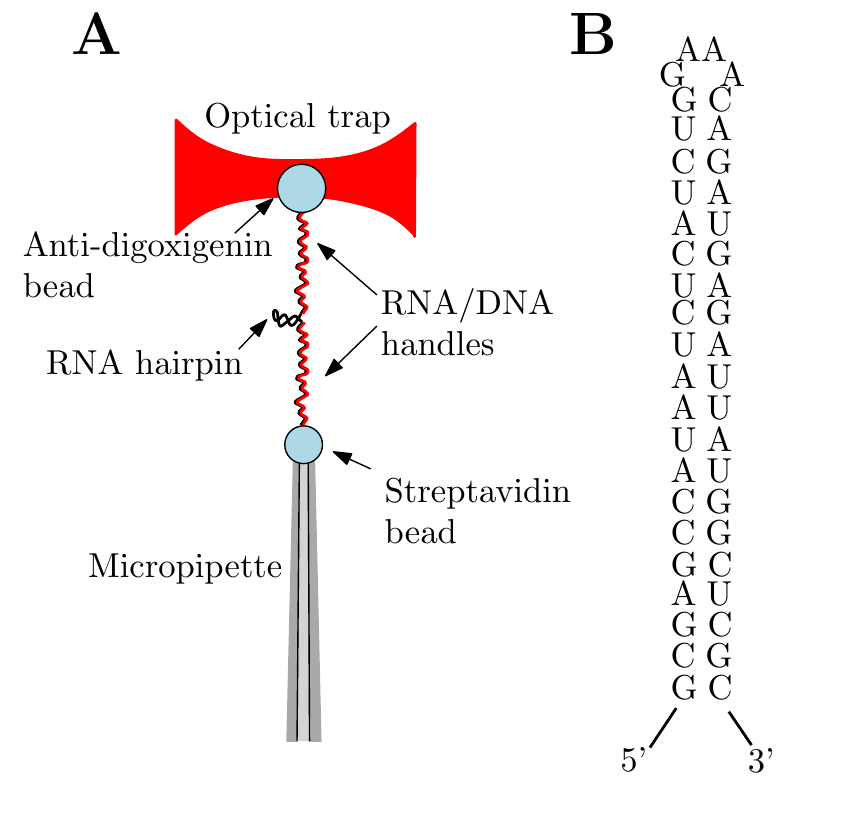}

\textbf{Figure 1. Experimental setup and RNA sequence.} (\textbf{A})
A single RNA hairpin is attached to two polystyrene microspheres through
RNA/DNA heteroduplexes used as handles. The anti-digoxigenin antibody-coated
microsphere is optically trapped while the streptavidin-coated microsphere
is positioned at the tip of a micropipette by air suction. (\textbf{B})
RNA hairpin sequence.

\clearpage{}
\includegraphics[scale=1]{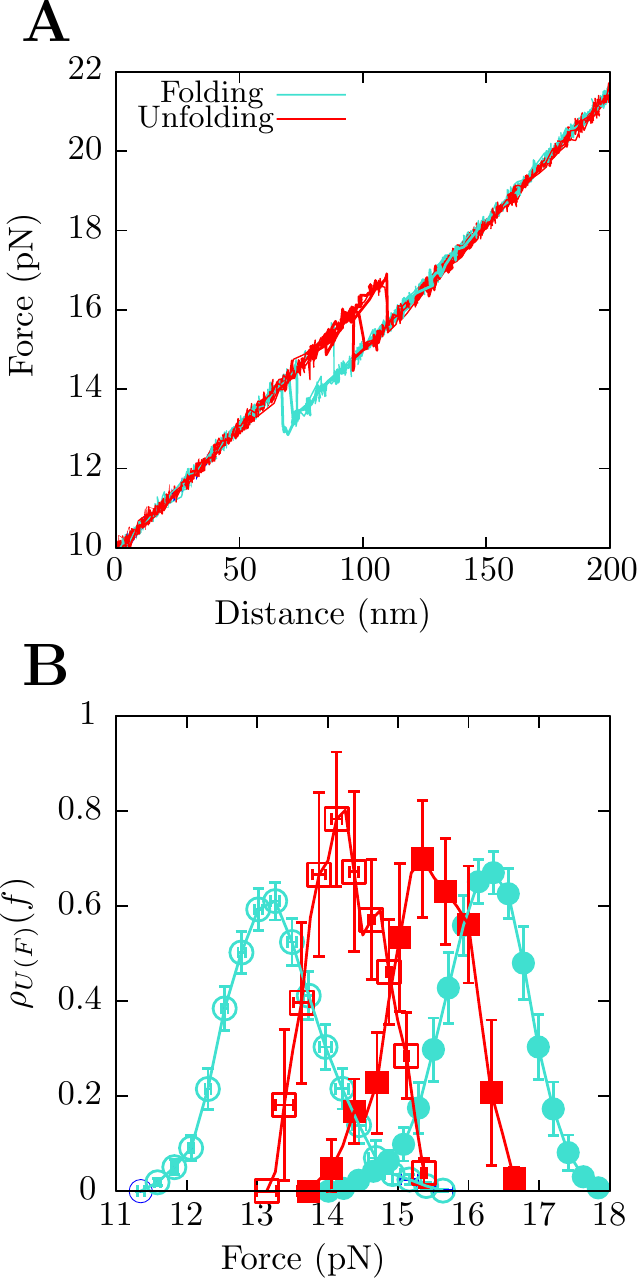}

\textbf{Figure 2. Pulling experiments.} (\textbf{A}) A few pulling
cycles for the RNA hairpin showing the unfolding (red) and folding
(blue) trajectories. (\textbf{B}) Experimental distribution for the
unfolding first rupture forces at 1.8 pN/s (red full squares) and 12.5 pN/s
(blue full circles), and for the folding first rupture forces at 1.8 pN/s (red empty squares)
and 12.5 pN/s (blue empty circles).

\clearpage{}
\includegraphics[scale=1]{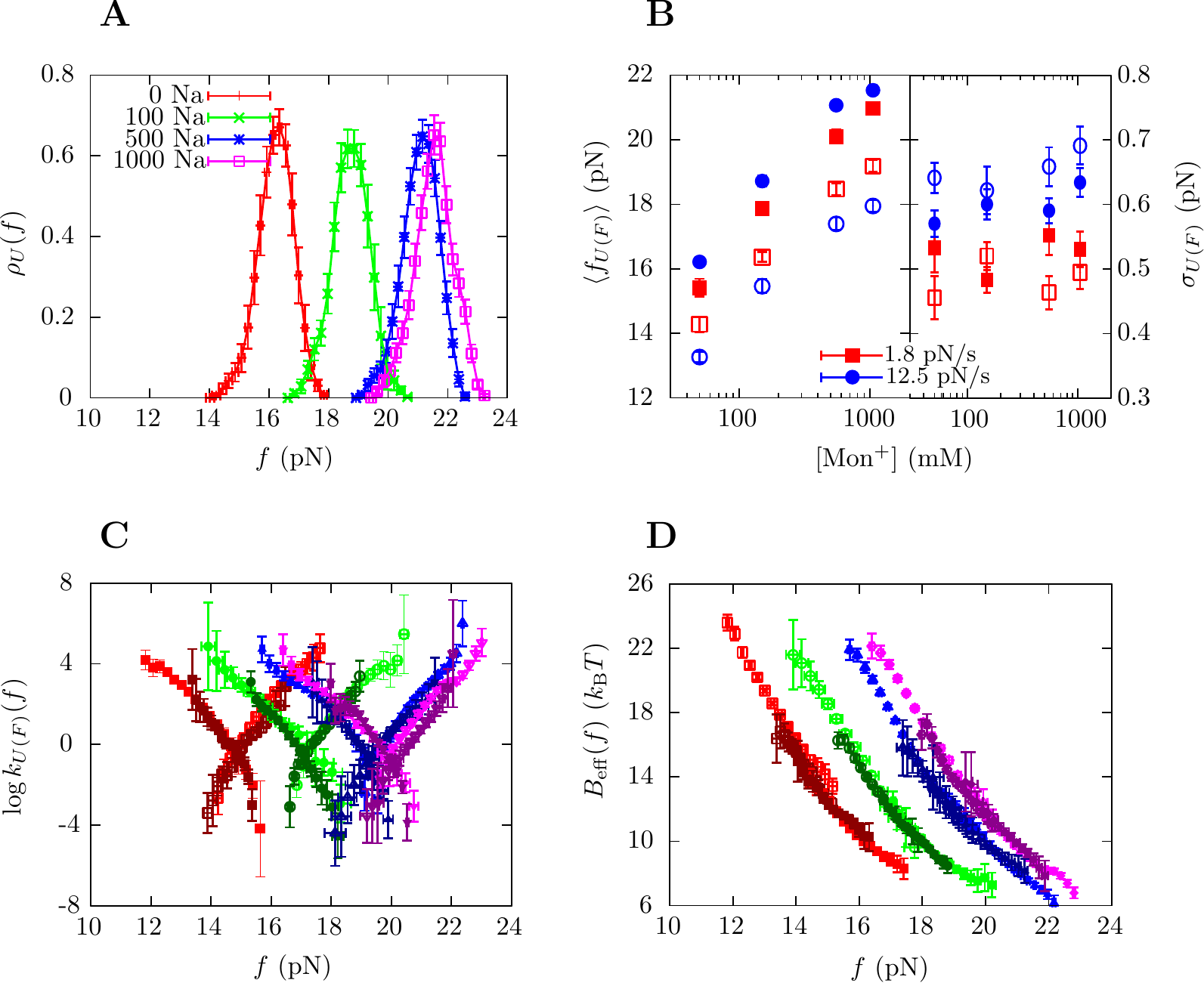}

\textbf{Figure 3. Kinetic analysis of experiments at varying [Mon$^{+}$].}
(\textbf{A}) Experimental distribution of the unfolding rupture forces
in buffers containing 50 mM (red), 150 mM (green), 550 mM (blue),
and 1050 mM (magenta) [Mon$^{+}$]. These experiments were done
at a loading rate of 1.8 pN/s. (\textbf{B}) Average rupture forces
$\left\langle f_{U(F)}\right\rangle $ and standard deviations $\sigma_{U(F)}$
as a function of monovalent cation concentration at loading rates
of 1.8 pN/s (red) and 12.5 pN/s (blue). Full symbols refer to unfolding
and empty symbols to folding. (\textbf{C}) Log-Linear plot of the
transition rates versus force. Experiments were done at 50 mM [Mon$^{+}$]
for loading rates of 1.8 pN/s (dark red) and 12.5 pN/s (red), at 150
mM [Mon$^{+}$] for loading rates of 1.8 pN/s (dark green) and
12.5 pN/s (green), at 550 mM [Mon$^{+}$] for loading rates of
1.8 pN/s (dark blue) and 12.5 pN/s (blue), and at 1050 mM [Mon$^{+}$]
for loading rates of 1.8 pN/s (dark violet) and 12.5 pN/s (magenta).
(\textbf{D}) Dependence of the effective barrier $B_{\rm eff}(f)$
on force at different [Mon$^{+}$]. Color code as in (\textbf{C}).
 
\clearpage{}
\includegraphics[scale=1]{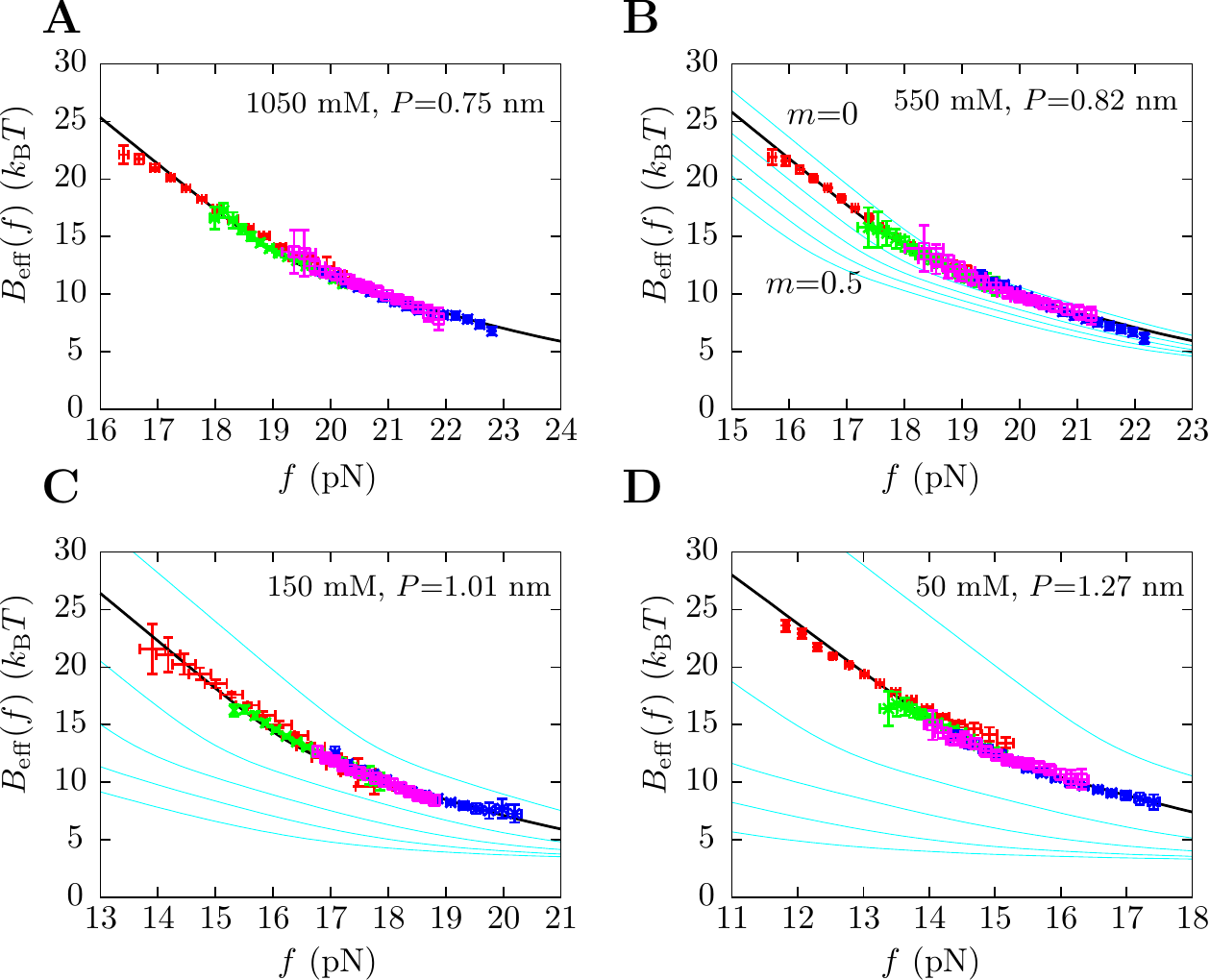}

\textbf{Figure 4. Determination of the salt correction parameter
\textit{\textbf{m}}}. Experimental evaluation of the sequence-independent correction parameter $m$ from $g_{1}([\text{Mon}^{+}])=m\log([\text{Mon}^{+}]/1000)$
using eq. \ref{eq: barrier_method}. Estimators of $B_{\rm eff}^{(U/F)}(f)$
obtained experimentally (eqs. \ref{eq: barrier_method}) were compared with the expected $B_{\rm eff}^{KT}(f)$ (eq. \ref{eq: theor_effBarrier}) 
profiles for different values of $m$ (0, 0.1, 0.2, 0.3, 0.4 and 0.5 from
top to bottom). %
Red (green) points are the experimental estimators $B_{\rm eff}^{(F)}(f)$
at a pulling rate of 12.5 (1.8) pN/s. Blue (magenta) points are the
experimental estimators of $B_{\rm eff}^{(U)}(f)$ at a pulling
rate of 12.5 (1.8) pN/s. Light blue lines are the $B_{\rm eff}^{KT}(f)$
profiles for values of $m$ not matched, and black lines are the experimental
estimators of $B_{\rm eff}(f)$ that match with experiments. Application
of the method to experiments done at 1050 mM $[\text{Mon}^{+}]$ %
(\textbf{A}), 550 mM $[\text{Mon}^{+}]$ (\textbf{B}), 150 mM $[\text{Mon}^{+}]$
(\textbf{C}), and 50 mM $[\text{Mon}^{+}]$ (\textbf{D}). 

\clearpage{}
\includegraphics[scale=1]{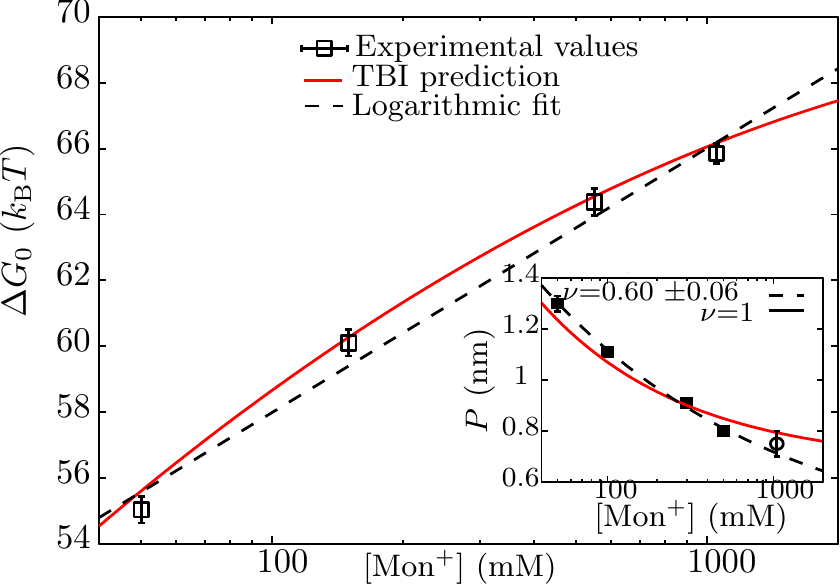}

\textbf{Figure 5. Free energy of formation of the RNA hairpin as
a function of [Mon$^{+}$].} Main panel: Free energy obtained experimentally
(squares), using the logarithmic dependence with salt concentration given by $g_1([{\rm Mon}^+])$ (dashed line) and using the TBI model (continuous line) Inset: Persistence
lengths $P$ obtained from the application of the Thick Chain model
to published experimental data for poly-U RNA stretching in buffers
containing 5, 10, 50, 100, 300, and 500 mM of [Na$^{+}$] (squares)
\cite{Seol2004,Toan2006}. We have included the value of the persistence
length $P$ that we obtained in this study at 1050 mM [Mon$^{+}$]
(empty circle). Two different fits to data were done from eq. \ref{eq: Persistence_length_Debye}:
a fixed value $\nu=1$ (red) and $\nu$ as free parameter $\nu=0.60\pm0.06$
(blue). 

\clearpage{}
\includegraphics[scale=1]{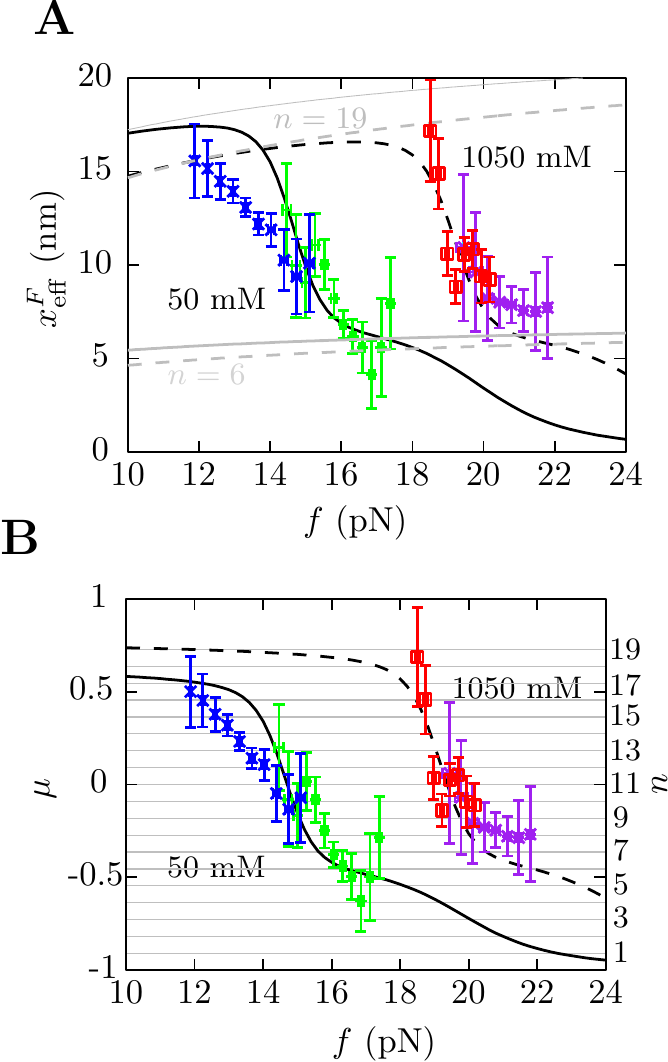}

\textbf{Figure 6. Barrier location and mechanical fragility at 50
mM and 1050 mM [Mon$^{+}$].} (\textbf{A}) Force-dependence of the barrier
position measured with respect to the F state, $x_{\rm eff}^{F}(f)$. Continuous gray line is the WLC prediction of the
molecular extension when $n=19$ or $n=6$ bps are unzipped 
at 50 mM [Mon$^{+}$], and dashed gray line corresponds
to the WLC prediction when $n=19$ or $n=6$ bps are unzipped at 1050 mM [Mon$^{+}$]. 
As seen, at an intermediate
value of forces $n=6$ coincides with the TS for both
ionic conditions. (\textbf{B}) Dependence of fragility $\mu(f)$ with
force. Gray lines indicate the value of the fragility for different
locations $n$ of the TS along the stem. Continuous black lines are
the theoretical prediction using Kramers rate theory for data at 50
mM [Mon$^{+}$], and dashed black lines for data at 1050 mM [Mon$^{+}$].
Blue and green points are the experimental evaluation of $x_{\rm eff}^{F}(f)$
and $\mu(f)$ for folding and unfolding data collected at 50 mM [Mon$^{+}$].
Red and purple points are the experimental evaluation for folding
and unfolding at 1050 mM [Mon$^{+}$].

\clearpage{}
\includegraphics[scale=1]{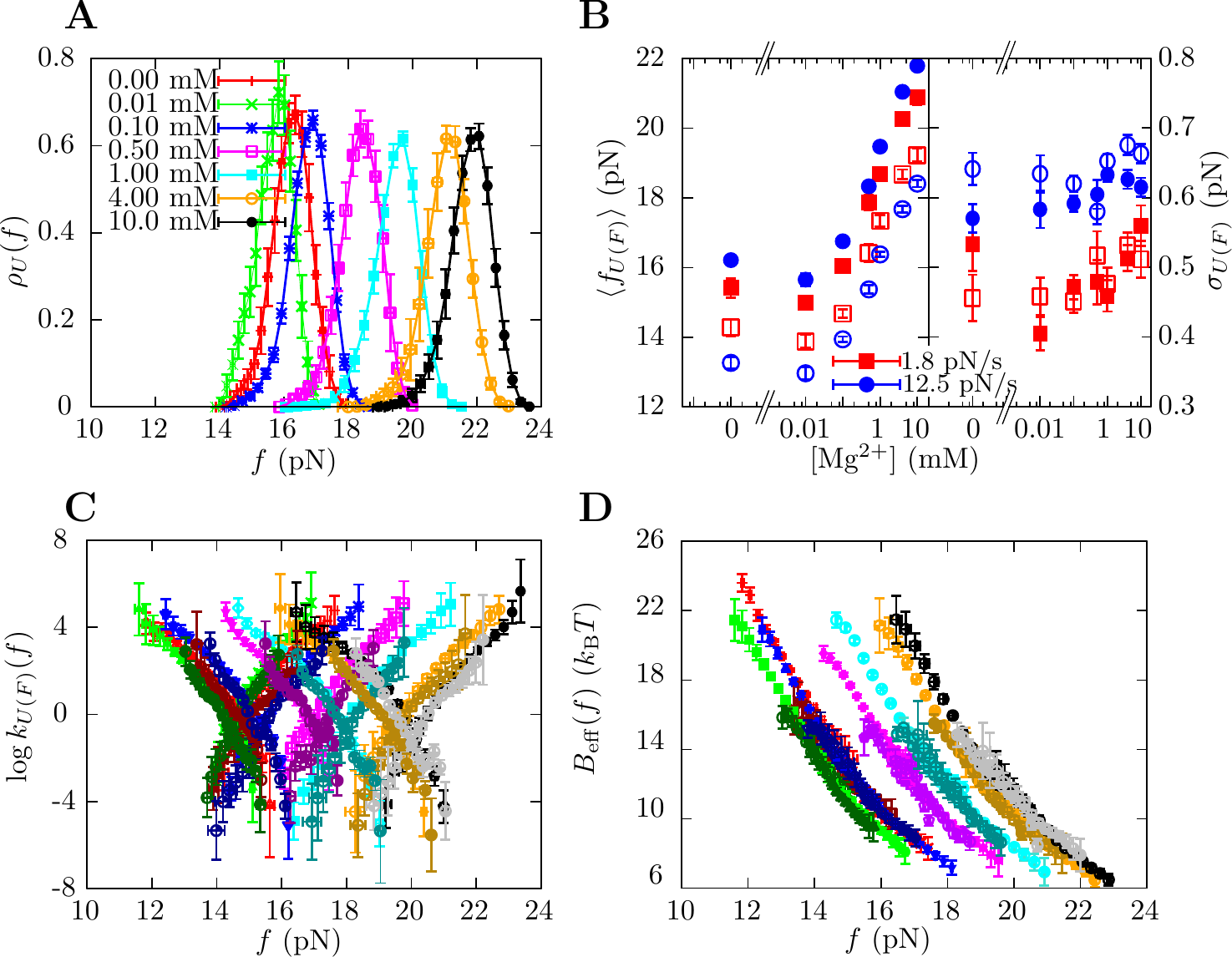}

\textbf{Figure 7. Kinetic analysis of experiments at varying [Mg$^{2+}$].}
(\textbf{A}) Experimental distribution of the unfolding rupture forces
in buffers containing 0.00 mM (red), 0.01 mM (green), 0.1 mM (blue),
0.5 mM (magenta), 1 mM (cyan), 4 mM (orange), and 10 mM (black) of
[Mg$^{2+}$] and 50 mM of monovalent cations. These experiments were
done at a loading rate of 1.8 pN/s. (\textbf{B}) Average rupture forces
$\left\langle f_{U(F)}\right\rangle $ and standard deviations $\sigma_{U(F)}$
obtained in experiments done at different [Mg$^{2+}$] and at
loading rates of 1.8 pN/s (red) and 12.5 pN/s (blue). Full symbols
refer to unfolding and empty symbols to folding. (\textbf{C}) Log-Linear
plot of the transition rates versus force. Experiments were done at
0.00 mM [Mg$^{2+}$] for loading rates of 1.8 pN/s (dark red)
and 12.5 pN/s (red), at 0.01 mM [Mg$^{2+}$] for loading rates
of 1.8 pN/s (dark green) and 12.5 pN/s (green), at 0.1 mM [Mg$^{2+}$]
for loading rates of 1.8 pN/s (dark blue) and 12.5 pN/s (blue), at
0.5 mM [Mg$^{2+}$] for loading rates of 1.8 pN/s (dark violet)
and 12.5 pN/s (magenta), at 1 mM [Mg$^{2+}$] for loading rates
of 1.8 pN/s (dark cyan) and 12.5 pN/s (cyan), at 4 mM [Mg$^{2+}$]
for loading rates of 1.8 pN/s (dark orange) and 12.5 pN/s (orange),
and at 10 mM [Mg$^{2+}$] for loading rates of 1.8 pN/s (gray)
and 12.5 pN/s (black). (\textbf{D}) Dependence of the effective barrier
$B_{\rm eff}(f)$ on force at different [Mg$^{2+}$]. Color
code as in (\textbf{C}).

\clearpage{}
\includegraphics[scale=1]{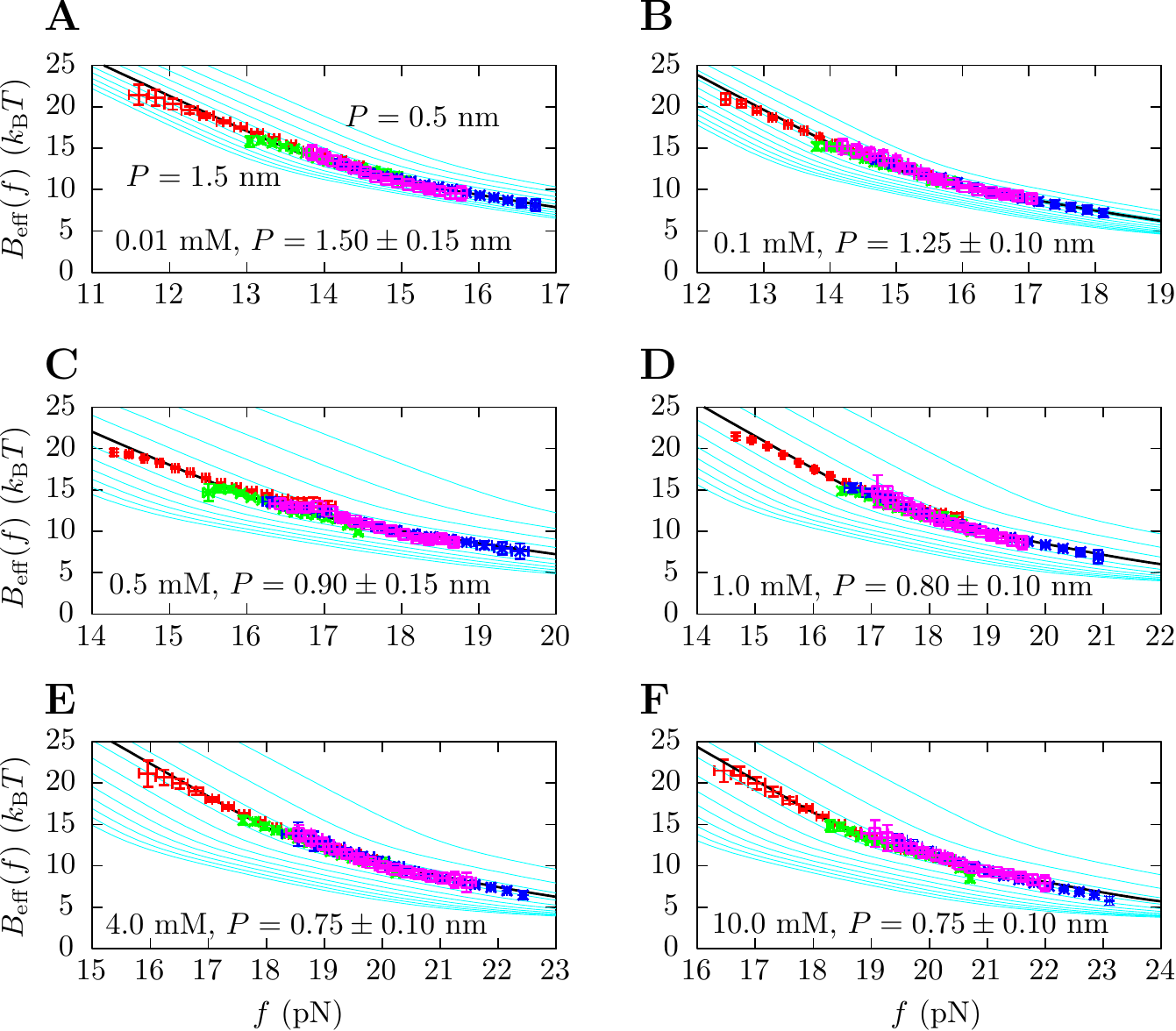}

\textbf{Figure 8. Determination of the persistence length of ssRNA
at varying [Mg$^{2+}$].} Estimators of $B_{\rm eff}^{(U/F)}(f)$
obtained experimentally were compared with the expected $B_{\rm eff}^{KT}(f)$
profiles for different values of $P$ (0.5, 0.6, 0.7, 0.8, 0.9, 1.0,
1.1, 1.2, 1.3, 1.4 and 1.5 nm from top to bottom) using eq. \ref{eq: theor_effBarrier} and eqs. \ref{eq: barrier_method}.
Red (green) points are the experimental estimators $B_{\rm eff}^{(F)}(f)$
at a pulling rate of 12.5 (1.8) pN/s. Blue (magenta) points are the
experimental estimators of $B_{\rm eff}^{(U)}(f)$ at a pulling
rate of 12.5 (1.8) pN/s. Light blue lines are the $B_{\rm eff}^{KT}(f)$
profiles for values of $m$ not matched, and black lines are the experimental
estimators of $B_{\rm eff}(f)$ that match the experiments. Application
of the method for experiments done at 0.01 mM [Mg$^{2+}$] (\textbf{A}),
0.1 mM [Mg$^{2+}$] (\textbf{B}), 0.5 mM [Mg$^{2+}$] (\textbf{C}),
1 mM [Mg$^{2+}$] (\textbf{D}), 4 mM [Mg$^{2+}$] (\textbf{E}),
and 10 mM [Mg$^{2+}$] (\textbf{F}). 

\clearpage{}
\includegraphics[scale=1]{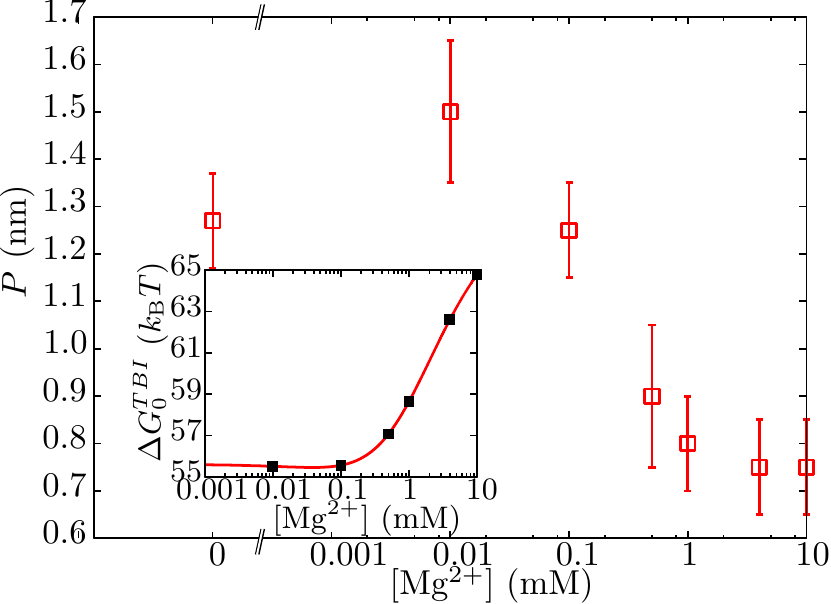}

\textbf{Figure 9. Dependence of the persistence length $P$ on [Mg$^{2+}$].}
Main panel: Experimental persistence length versus [Mg$^{2+}$].
Inset: Dependence of the free energy of formation of the RNA hairpin
$\Delta G_{\rm N}^{\text{TBI}}(0)$ on [Mg$^{2+}$] with fixed
50 mM [Mon$^{+}$] obtained using the TBI model \cite{Tan2007,Tan2008}.
Black points are the values of the free energy of formation that we
used for our analysis.

\clearpage{}
\includegraphics[scale=1]{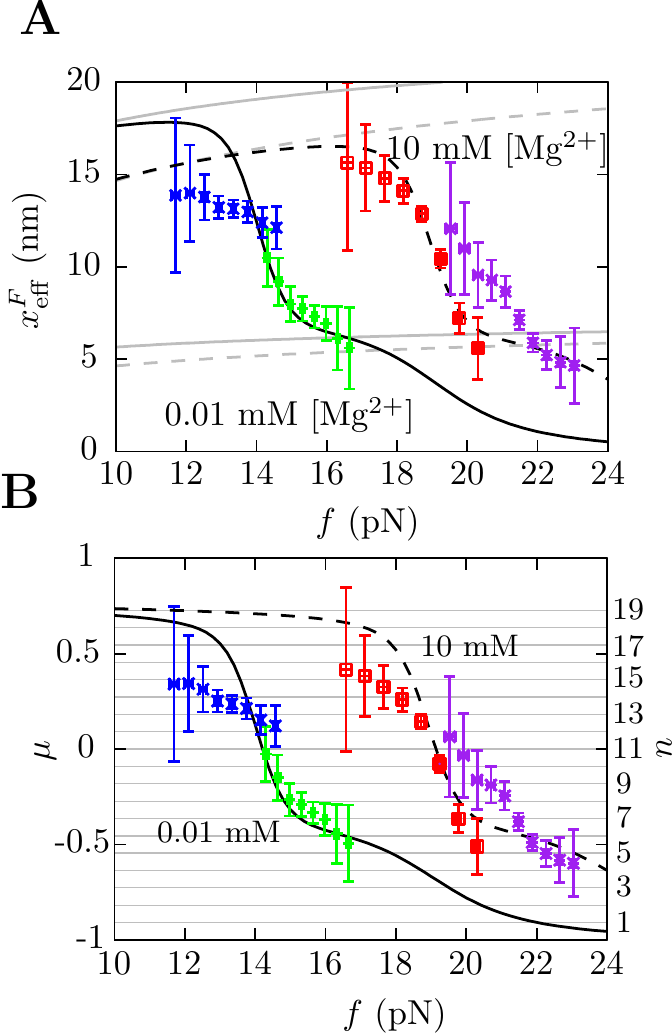}

\textbf{Figure 10. Barrier location and mechanical fragility at 0.01
mM and 10 mM [Mg$^{2+}$].} (\textbf{A}) Force-dependence of the barrier
position measured with respect to the F state, $x_{\rm eff}^{F}(f)$.
Continuous gray line is the WLC prediction of the molecular
extension when $n=19$ or $n=6$ bps are unzipped at 0.01 mM
[Mg$^{2+}$], and dashed gray line corresponds to the
WLC prediction when $n=19$ or $n=6$ bps are unzipped 
at 10 mM [Mg$^{2+}$]. At an intermediate range of force the
TS coincides with $n=6$ for both ionic conditions. %
(\textbf{B}) Dependence of fragility $\mu(f)$ at 0.01 mM and 10 mM
[Mg$^{2+}$]. Gray lines indicate the value of the fragility for
different locations $n$ of the TS along the stem. Continuous black
lines are the theoretical prediction using Kramers rate theory for
data at 0.01 mM [Mg$^{2+}$], and dashed black lines for data
at 10 mM [Mg$^{2+}$]. Blue and green points are the experimental
evaluation of $x_{\rm eff}^{F}(f)$ and $\mu(f)$ for folding and
unfolding data collected at 0.01 mM [Mg$^{2+}$]. Red and purple
points are the experimental evaluation for folding and unfolding at
10 mM [Mg$^{2+}$]. 

\clearpage{}
\includegraphics[scale=1]{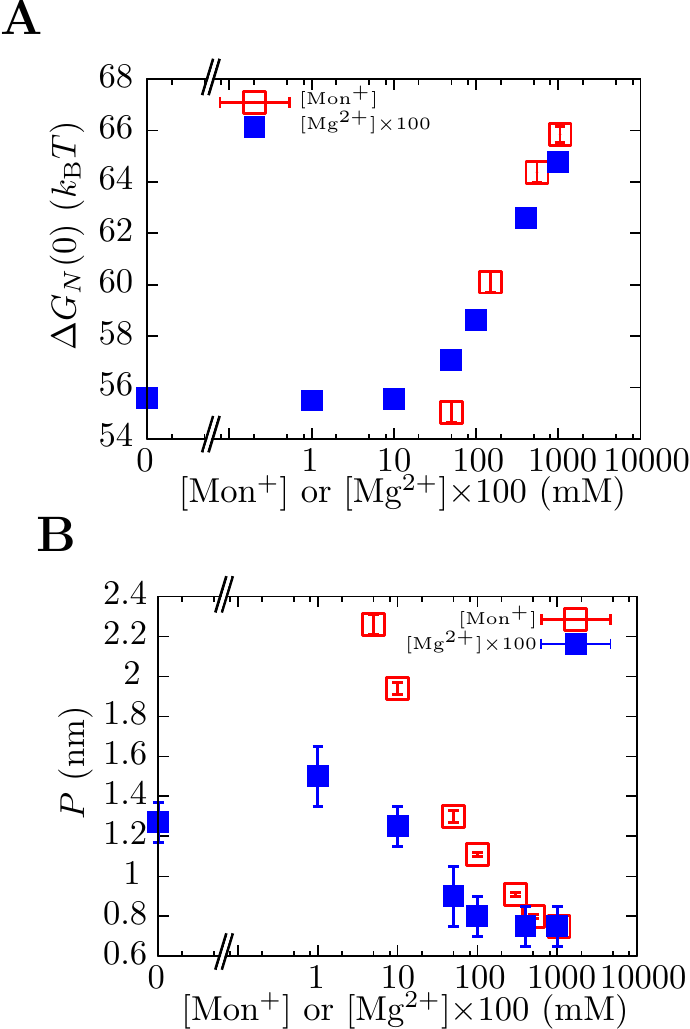}

\textbf{Figure 11. Comparison between [Mon$^{+}$] and [Mg$^{2+}$]
results.} (\textbf{A}) Free energy of formation of the RNA hairpin
at different salt conditions. Magnesium concentrations have been multiplied
by 100 along the horizontal axis. %
(\textbf{B}) Persistence length values for the ssRNA hairpin at
different salt conditions. Magnesium concentrations have been multiplied
by 100 along the horizontal axis. 

\newpage{}

\section{Tables}

\begin{table}[ht]
\centering \begin{tabular}{cr@{$\pm$}lr@{$\pm$}lr@{$\pm$}l}
\hline 
[Mon$^{+}$] (mM)  & \multicolumn{2}{c}{$P$ (nm)} & \multicolumn{2}{c}{$\log k_{0}$ (1/s)} & \multicolumn{2}{c}{$\Delta G_{\rm N}(0)$ ($k_{\rm B}T$)}\\
\hline 
1050  & 0.75  & 0.05  & 10.9  & 0.4  & 65.1  & 0.3 \\
550  & 0.82  & 0.02  & 10.5  & 0.4  & 64.0  & 0.4 \\
150  & 1.01  & 0.01  & 11.3  & 0.5  & 59.4  & 0.4 \\
50  & 1.27  & 0.03  & 12.4  & 0.4  & 54.0  & 0.4 \\
\hline
\end{tabular}\caption{\textbf{ Parameters obtained from experiments at different [Mon$^{+}$].}
ssRNA persistence length $P$, $\log k_{0}$, and free energy of formation
($\Delta G_{\rm N}(0)$) for the RNA hairpin at different monovalent
ion concentrations.}
\label{tab: tab1} 
\end{table}


\begin{table}[ht]
\centering \begin{tabular}{cr@{$\pm$}lr@{$\pm$}lcr@{$\pm$}l}
\hline 
[Mg$^{2+}$] (mM)  & \multicolumn{2}{c}{$P$ (nm)} & \multicolumn{2}{c}{$\log k_{0}$ (1/s)} & $\Delta G_{\rm N}^{TBI}(0)$ ($k_{\rm B}T$) & \multicolumn{2}{c}{$g_{2}$ (kcal/mol)}\\
\hline 
0.00  & 1.27  & 0.03  & 12.40  & 0.40  & 55.58  & 0.000  & 0.005\\
0.01  & 1.50  & 0.15  & 12.05  & 0.30  & 55.50  & -0.007  & 0.005\\
0.10  & 1.25  & 0.10  & 11.45  & 0.30  & 55.55  & -0.005  & 0.005\\
0.50  & 0.90  & 0.15  & 11.50  & 0.30  & 57.06  & 0.0393  & 0.005\\
1.00  & 0.80  & 0.10  & 11.40  & 0.50  & 58.63  & 0.0858  & 0.005\\
4.00  & 0.75  & 0.10  & 11.15  & 0.50  & 62.60  & 0.2033  & 0.005\\
10.0  & 0.75  & 0.10  & 10.40  & 0.50  & 64.77  & 0.2678  & 0.005\\
\hline
\end{tabular}\caption{\textbf{Parameters obtained for experiments at different [Mg$^{2+}$].}
Persistence length for ssRNA $P$, $\log k_{0}$, theoretical predictions
for the free energies of formation based on the TBI model, $\Delta G_{\rm N}^{TBI}(0)$,
and sequence-independent correction $g_{2}([$Mg$^{2+}])$ for the RNA hairpin
at different magnesium concentrations. A fixed concentration of 50
mM [Mon$^{+}$] was used in all ionic conditions. }
\label{tab: tab2} 
\end{table}

\newpage

\section*{Supporting Material}

\section{Molecules studied}

\begin{table}[ht]
\centering
\begin{tabular}{ccccc}
\hline
[Mon$^+$] (mM) & [Mg$^{2+}$] (mM) & loading rate (pN/s) & Molecules &  Total cycles \\
\hline
50   & 0 & 1.8  & 4 & 95  \\
50   & 0 & 12.5 & 2 & 376 \\
150  & 0 & 1.8  & 6 & 292 \\ 
150  & 0 & 12.5 & 4 & 329 \\ 
550  & 0 & 1.8  & 5 & 163 \\
550  & 0 & 12.5 & 3 & 501 \\
1050 & 0 & 1.8  & 9 & 185 \\ 
1050 & 0 & 12.5 & 9 & 405 \\ 
50 & 0.01 & 1.8  & 5 & 146  \\
50 & 0.01 & 12.5 & 5 & 386  \\
50 & 0.1  & 1.8  & 7 & 374  \\
50 & 0.1  & 12.5 & 9 &  1434\\
50 & 0.5  & 1.8  & 2 & 112  \\
50 & 0.5  & 12.5 & 2 & 533  \\ 
50 & 1.0  & 1.8  & 4 & 205  \\ 
50 & 1.0  & 12.5 & 6 & 2183 \\
50 & 4.0  & 1.8  & 7 & 385  \\
50 & 4.0  & 12.5 & 7 & 1112 \\
50 & 10.0 & 1.8  & 7 & 190  \\
50 & 10.0 & 12.5 & 3 & 1189 \\
\hline
\end{tabular}
\caption{Number of molecules and total cycles measured at each ionic salt condition.}
\end{table}

\clearpage
              
\section{Study of fraying}

The phenomenon of ``fraying'' at the ends of DNA
and RNA duplexes can potentially interfere
with both solution and single-molecule measurements of DNA and RNA
stabilities, and it was previously suggested that its effects should
be introduced in data analysis \cite{Woodside2006}.
In order to check if fraying has an important role in our sequence, 
we computed the released (absorbed) molecular extension $\Delta x_m$ in
 the unfolding (folding) process. This can be done using the expression:
\begin{equation}
\label{eq: Dx exp}
\Delta x_m = \frac{\Delta f}{k_{\rm eff}}
\end{equation}
where $\Delta f$ is the force jump measured along the force-distance curve
(FDC) and $k_{\rm eff}$ is its slope before the transition. The change in
the molecular extension is also equal to:
\begin{equation}
\label{eq: dx th}
\Delta x_m=x_N(f)-x_n(f)
\end{equation}
where $x_N(f)$ is the equilibrium end-to-end distance of the 
unzipped hairpin evaluated at the unfolding/folding force ($N=20$ in this case); $x_n(f)$ is the projection of the folded hairpin along 
the force axis; and $n$ denotes the number of open/frayed base pairs in the F state. Ideally, in the absence of fraying, $n=0$. However, in presence of fraying we should find the value $n>0$ such that eq. \eqref{eq: Dx exp} and \eqref{eq: dx th} give the same change in molecular extension.

From the FDC we obtain $\Delta f=1.2\pm0.1$ pN and $k_{\rm eff}=0.0625\pm0.0146$ pN/nm 
which, using eq. \eqref{eq: Dx exp}, gives $\Delta x_m=19\pm2$ nm for any salt concentration. On the 
other hand, we evaluate $x_N(f)$ and $x_{n=0}(f)$ using the elastic 
properties summarized in Tables 1 and 2 at the measured unfolding/folding forces. 
Using eq. \eqref{eq: dx th} for $n=0$ we obtain that predicted values
for $\Delta x_m$ lie in the range between 18.2 and 19.7 nm. Therefore, we conclude that
fraying is not important for the molecule under study because the experimental evaluation of $\Delta x_m$ (eq. \eqref{eq: Dx exp}) is in agreement with the estimation of the released molecular extension for $N=20$ and $n=0$.

The effect of fraying has been proved to play an important role in former single-molecule stretching experiments, like in Woodside \textit{et. al.} \cite{Woodside2006}. 
To understand when fraying is relevant in RNA or DNA hairpins we can take a look to sequences at the beginning of the stem as a higher GC-content makes the structure more stable. 
For instance, the RNA hairpin studied here starts 
with 5'-GCG-3', whereas most sequences studied by Woodside \textit{et al}. \cite{Woodside2006} start with 5'-GAG-3'
(except two sequences that start with 5'-TAT-3' and 5'-AAG-3'). 
In Table \ref{tab: fraying} we compute the free energy difference $\Delta G_1(f)$ at 1 M NaCl between the completely 
folded conformation ($n=0$) and the frayed configuration with one open distal 
base pair ($n=1$) at different values of force for our RNA hairpin and molecule 20R55/4T in \cite{Woodside2006}.
At zero 
force, $\Delta G_{1}^{20R55/4T}(0)$ is below 3 $k_{\rm B}T$ and consequently thermal fluctuations
can overcome the energetic barrier and the frayed conformation can take place. However, in the case of our hairpin 
$\Delta G_1$ is too high for thermal fluctuations to overcome the energetic barrier. The same 
trend is observed at 10 pN. At 20 pN fraying is irrelevant because both molecules are in the unfolded state.

\begin{table}[ht]
 \centering
\begin{tabular}{|c|c|c|}
\hline 
Force (pN) & $\Delta G_{1}^{20R55/4T}$ $(k_{\rm B}T)$ & $\Delta G_{1}^{RNA}$ $(k_{\rm B}T)$\\
\hline
\hline 
0 & 2.67 & 6.45\\
\hline 
10 & 1.08 & 4.86\\
\hline 
20 & -1.34 & 2.45\\
\hline
\end{tabular}
\caption{Free energy differences between frayed
($n=1$) and completely closed structures ($n=0$) of 20R55/4T hairpin \cite{Woodside2006} and our RNA hairpin at 1 M NaCl and different
forces.}\label{tab: fraying}
\end{table}

Motivated by solution measurements \cite{Cain1997,Sarkar2010}, we also considered
the possibility of fraying at the opposite end of the stem,
on the base pairs closest to the loop. However, we found free energy differences of more than 10
$k_{\rm B}T$ between these configurations and the closed configuration, and
consequently we conclude that these ``frayed'' configurations
are not affecting our results.

Based on these considerations, we conclude that fraying
plays a rather minor role (if any) on the thermodynamics and kinetics
of folding/unfolding of the RNA hairpin and can be neglected.
              

\section{Derivation of the effective barrier $\mathbf{B_{\rm\bf eff}^{KT}(f)}$}
\label{sec: kramers}

Here we derive the analytical expression for the effective barrier of a one-dimensional free energy landscape based on the work by Hyeon and Thirumalai \cite{Hyeon2007}.

\begin{figure}[ht]
\centering
\includegraphics[scale=1]{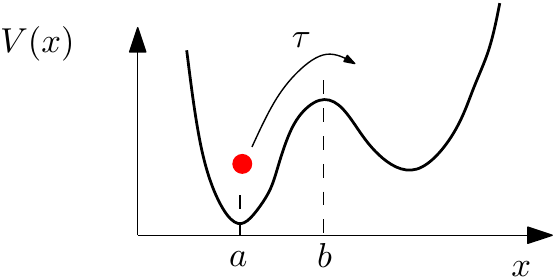}
\caption{Brownian particle in a double well potential.}\label{fig: FK}
\end{figure}

Suppose the system sketched in Fig. \ref{fig: FK}, where a Brownian particle is subject to the one dimensional potential $V(x)$. The time evolution of the probability density function (pdf) $p(x,t)$ to find the particle at the position $x$ at time $t$ follows the Fokker-Planck equation \cite{Hyeon2007,Zwanzig2001}:

\begin{align}
 \label{eq: FokkerPlanck}
\frac{\partial p(x,t)}{\partial t} &= D\frac{\partial}{\partial x}\left[\frac{\partial}{\partial x}+\frac{1}{k_{\rm B}T}\frac{dV(x)}{dx}\right] p(x,t)\nonumber\\
&= D\frac{\partial}{\partial x}\left[e^{-\frac{V(x)}{k_{\rm B}T}}\frac{\partial}{\partial x}e^{\frac{V(x)}{k_{\rm B}T}}\right]p(x,t)\nonumber\\
&=\mathcal{L}_{\rm FP}\left[p(x,t)\right]
\end{align}

Where $D$ is the diffusion coefficient and $\mathcal{L}_{FP}$ is the Fokker-Plank operator. 
If we suppose that the particle initially is located at $x=a$, $p(x,0)=\delta(x-a)$, the formal solution of equation \eqref{eq: FokkerPlanck} is:
\begin{equation}
p(x,t)=e^{t\mathcal{L}_{FP}}\delta(x-a)
\end{equation}

We want to evaluate the average time $\tau$ it takes to the Brownian particle to jump the kinetic barrier located at $x=b$. In order to simplify the following calculations, we suppose that there are absorbing conditions at $x=b$: once the particle reaches the maximum sketched in Fig. \ref{fig: FK} it always goes to the right well. The probability to find the particle in the left well of the potential $V(x)$ ($x\in[-\infty,b]$) at time $t$, also known as the survival probability $S(t)$, can be defined as:

\begin{align}
S(t)=\int_{-\infty}^b dx \text{ } p(x,t)
\end{align}

The time derivative of the survival probability is equal to the time survival pdf $\rho(t)$, that is, the pdf of the time it takes to the Brownian particle to cross the barrier located at $x=b$. 

\begin{align}
S(t+dt)-S(t)=-\rho(t)dt\Rightarrow \rho(t)=-\frac{\partial S(t)}{\partial t}
\end{align}

Therefore, the mean first passage time $\tau$ can be calculated as:

\begin{align}
\label{eq: tau}
\tau &= \int_0^\infty dt \text{ }t \text{ }\rho(b,t)\nonumber\\
&=\int_0^\infty dt ~ S(b,t) \nonumber\\
&=\int_0^\infty dt\int_{-\infty}^b dx ~ p(x,t) \nonumber\\
&=\int_0^\infty dt\int_{-\infty}^b dx ~ e^{t\mathcal{L}_{FP}}\delta(x-a) \nonumber\\
&=\int_0^\infty dt\int_{-\infty}^b dx ~ \delta(x-a)e^{t\mathcal{L}_{FP}^\dag}1 \nonumber\\
&=\int_0^\infty dt ~ e^{t\mathcal{L}_{FP}^\dag}1
\end{align}
In order to obtain this expression we integrated by parts and used that the adjoint operator satisfies $f(x)\mathcal{L}\left[g(x)\right]=g(x)\mathcal{L}^\dag \left[f(x)\right]$. If we apply the adjoint Fokker-Plank operator $\mathcal{L}_{FP}^\dag$ at both sides of equation \eqref{eq: tau} we obtain:

\begin{align}
\mathcal{L}_{FP}^\dag\tau &= \int_0^\infty dt ~ \mathcal{L}_{FP}^\dag e^{t\mathcal{L}_{FP}^\dag}1\nonumber\\
&= \int_0^\infty dt ~ \frac{de^{t\mathcal{L}_{FP}^\dag}}{dt}1\nonumber\\
&= -1
\end{align}
Which gives a differential equation for the survival time $\tau$ that depends on the adjoint Fokker-Plank operator.

It can be demonstrated that:
\begin{equation}
\label{eq: adjoint}
\mathcal{L}^\dag_{FP}=De^{\frac{V(x)}{k_{\rm B}T}}\frac{\partial e^{-\frac{V(x)}{k_{\rm B}T}}}{\partial x} \frac{\partial}{\partial x}
\end{equation}
and we can write the following differential equation:
\begin{equation}
\label{eq: dif}
\mathcal{L}_{FP}^\dag \tau(x)= De^{\frac{V(x)}{k_{\rm B}T}}\frac{\partial}{\partial x}\left(e^{-\frac{V(x)}{k_{\rm B}T}}\right)\frac{\partial}{\partial x} \tau(x)=-1
\end{equation}

In order to solve eq. \eqref{eq: dif} we use the absorbing boundary condition $\tau(b)=0$ and the reflecting boundary $\frac{\partial \tau}{\partial x}|_{x=a}=0$. 
\begin{equation}
\tau(x)=\frac{1}{D}\int_x^b dy~e^{\frac{V(y)}{k_{\rm B}T}}\int_a^y dx e^{-\frac{V(x)}{k_{\rm B}T}}
\end{equation}

Once we have evaluated the average survival time of the Brownian particle, we want to evaluate the effective barrier of the potential $V(x)$. Using the phenomenological Arrhenius approach \cite{Arrhenius} that considers that the survival time depends on the exponential of the barrier we can write:
\begin{equation}
 \tau\simeq e^{\frac{B_{\rm eff}}{k_{\rm B}T}}\Rightarrow B_{\rm eff}=k_{\rm B}T\log\left( \tau /\tau_0\right)
\end{equation}
\begin{equation}
\label{eq: Bcontinua}
B_{\rm eff}= k_{\rm B}T\log\left( \frac{1}{\tau_0D}\int_x^b dy~e^{\frac{V(y)}{k_{\rm B}T}}\int_a^y dx e^{-\frac{V(x)}{k_{\rm B}T}}\right)
\end{equation}
$\tau_0$ is related to the diffusion time of the particle the along $x$ axis. By discretization of equation \eqref{eq: Bcontinua} we obtain the expression (7) of the main paper, 
\begin{equation}
\label{eq: theor_effBarrier2}
B_{\rm eff}(f)=k_{\rm B}T\log\left[\sum_{n=0}^{N}e^{\frac{\Delta G_{{\rm {n}}}(f)}{k_{\rm B}T}}\left(\sum\limits _{n'=0}^{n}e^{-\frac{\Delta G_{\rm n'}(f)}{k_{\rm B}T}}\right)\right]
\end{equation}
Where $\Delta G_n$ is the potential energy $V(x)$ and where we considered that $D\tau_0\simeq\mathcal{O}(1)$.

\clearpage

\section{Sensitivity of the data analysis}

An experimental estimation of the kinetic barrier is obtained from the measured transition rates $k_U(f)$ and $k_F(f)$. From the unfolding transition rate, the estimator of the kinetic barrier is given by:
\begin{equation}
\label{eq: BU}
\frac{B_{\rm eff}^{(U)}(f)}{k_{\rm B}T} = \log k_0-\log k_U(f)
\end{equation}
Where $\log k_0$ is a constant (equal to the logarithm of the attempt rate at zero force for the activated kinetics) and $\log k_U(f)$ is estimated from the measured unfolding rupture forces (see section 2.6 in the main paper).

On the other hand, from folding rupture forces the estimation of the kinetic barrier is:
\begin{align}
\label{eq: BF}
\frac{B_{\rm eff}^{(F)}(f)}{k_{\rm B}T} &= \log k_0-\log k_F(f) + \frac{\Delta G_{\rm N}(f)}{k_{\rm B}T} \nonumber\\
&=\log k_0 -\log k_F(f) + \frac{\Delta G_{\rm N}(0)}{k_{\rm B}T} + \frac{\Delta G_{\rm N}^{ssRNA}(f)}{k_{\rm B}T}+\frac{\Delta G_{\rm N}^{d_0}(f)}{k_{\rm B}T}\nonumber\\
&=\log k_0-\log k_F(f) + \frac{\Delta G_{\rm N}(0)}{k_{\rm B}T} - \frac{\int_0^f x_{\rm N}^{ssRNA}(f')df'}{k_{\rm B}T}+\log\left[\frac{k_{\rm B}T}{fd_0}\sinh\left(\frac{fd_0}{k_{\rm B}T}\right)\right]\nonumber\\
\end{align}
Where $\log k_0$ is the same constant as in eq. \eqref{eq: BU} and $\log k_F(f)$ is obtained from the measured folding rupture forces. The term $\Delta G_{\rm N}(0)/k_{\rm B}T$ is another constant equal to the free energy of the RNA hairpin at zero force; $\int_0^f x_{\rm N}^{ssRNA}(f')df'$ is a force dependent term evaluated according to the model used to describe the elastic response of ssRNA (here we use the WLC model with a salt-dependent persistence length $P$); and the last term, $\log\left[\frac{k_{\rm B}T}{fd_0}\sinh\left(\frac{fd_0}{k_{\rm B}T}\right)\right]$, is evaluated using $d_0=2.0$ nm at any salt condition for the given force $f$ (see section 2.3).

Typically, $\log k_0$ is unknown and either $\Delta G_N(0)/k_{\rm B}T$ is unknown and $P$ is known (here, for monovalent salt conditions) or  $\Delta G_N(0)/k_{\rm B}T$ is known and $P$ is unknown (for mixed monovalent/Mg$^{2+}$ conditions).

\subsection{Sensitivity of the method at determining $\mathbf{\Delta G_N(0)}$}
\label{sec: SDG}

\begin{figure}[ht]
\centering
\includegraphics[scale=1]{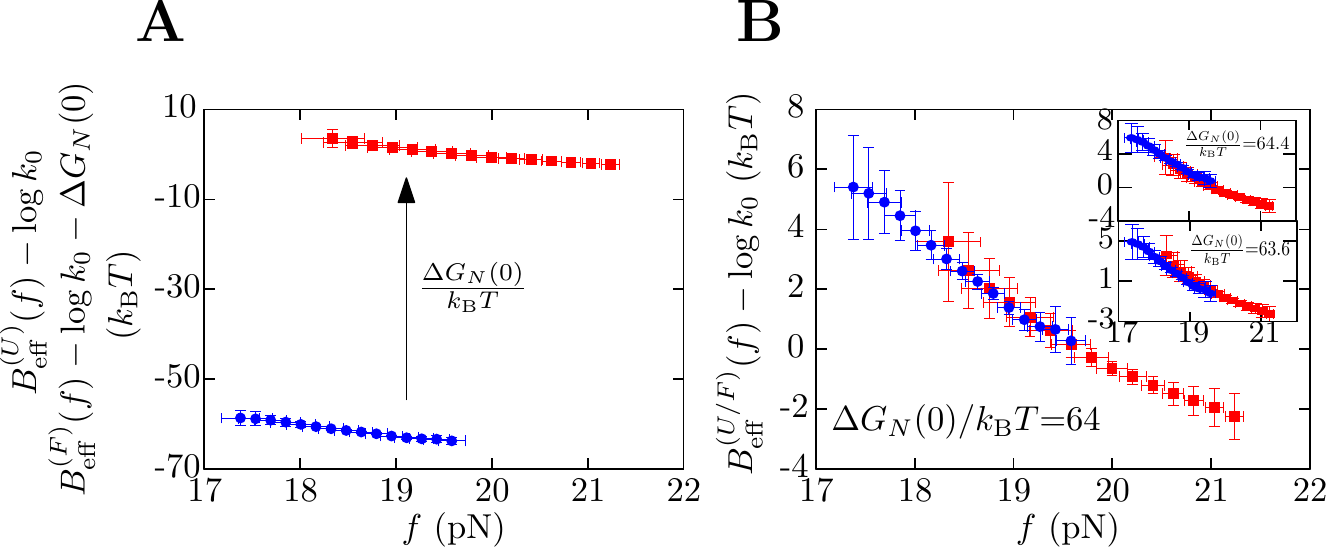}
\caption{\textbf{Determination of $\mathbf{\Delta G_N(0)}$.} Data obtained at 1.8 pN/s and 550 mM [Mon$^+$].
\textbf{(A)} Evaluation of $B^{(U)}(f)-\log k_0$ (red squares) and $B^{(F)}(f)-\log k_0-\Delta G_N(0)/k_{\rm B}T$ (blue circles).
\textbf{(B)} $\Delta G_N(0)/k_{\rm B}T$ is obtained using the overlapping of the kinetic barrier with force. Once unfolding and folding data overlap, the resulting experimental curve is equal to $B_{\rm eff}(f)-\log k_0$. \textit{Insets}: The value of $\Delta G_N(0)/k_{\rm B}T$ can be overestimated (top) or underestimated (bottom) so that the continuity requirement is not well-satisfied.
}\label{fig: analisi1}
\end{figure}

For a given value of the persistence length $P$ we can evaluate the kinetic barrier from experimental unfolding/folding rupture forces by ignoring the unknown constants ($\log k_0$ and $\Delta G_N(0)/k_{\rm B}T$) and we obtain the result shown in Fig. \ref{fig: analisi1}A. Error bars are evaluated using the bootstrap method. 

In order to determine the constant $\Delta G_N(0)/k_{\rm B}T$, \textit{i. e.} the free energy of formation of the RNA hairpin, we impose the continuity of the kinetic barrier in folding and unfolding data (Fig. \ref{fig: analisi1}B). The error committed in the evaluation of $\Delta G_N(0)/k_{\rm B}T$ mainly depends on the good agreement of the overlapping between $\left(B_{\rm eff}^{(U)}(f)-\log k_0\right)$ and $\left(B_{\rm eff}^{(F)}(f)-\log k_0\right)$ and their error bars. In the example provided in Fig. \ref{fig: analisi1}B the best overlapping is found at $\Delta G_N(0)=64$ $k_{\rm B}T$. In the insets we see that the continuity requirement worsens for values of $\Delta G_N(0)$ as close as 64.4 or 63.6 $k_{\rm B}T$. As a consequence, we estimate $\Delta G_N(0)=64.0\pm0.4$ $k_{\rm B}T$ (see Table 1 in the main document). 

\subsection{Sensitivity of the method at determining $\mathbf{\log k_0}$}

\begin{figure}[ht]
\centering
\includegraphics[scale=1]{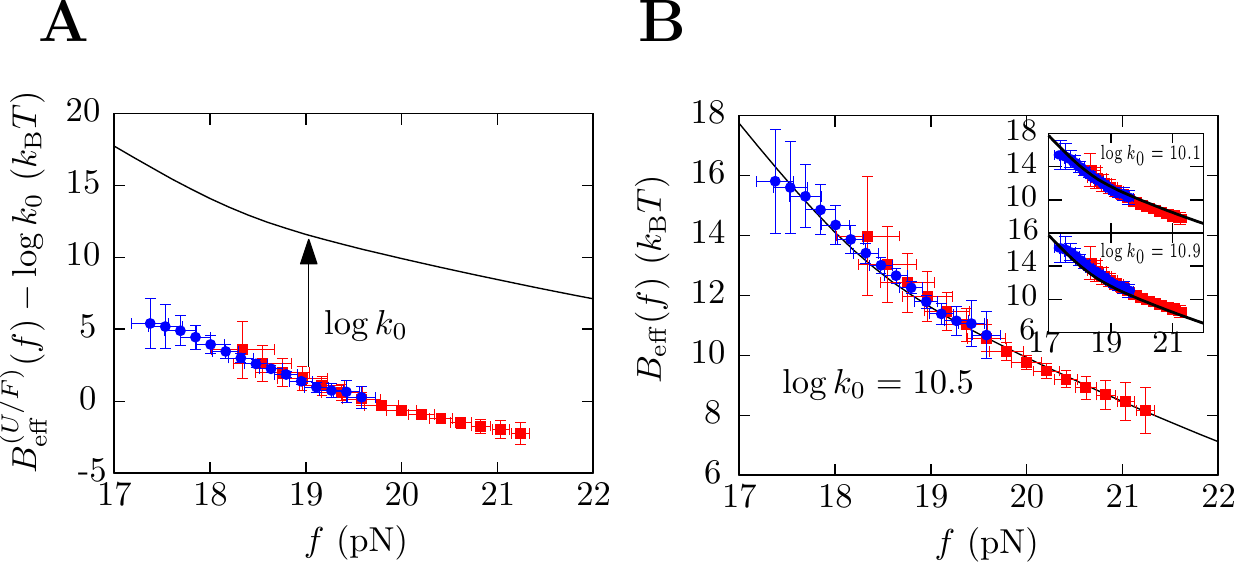}
\caption{\textbf{Determination of $\mathbf{\log k_0}$.} Data obtained at 1.8 pN/s and 550 mM [Mon$^+$].
\textbf{(A)} Black-straight line is the theoretical evaluation of the kinetic barrier using Kramers theory (section S3 and section 2.5 in the main document) and blue circles (red squares) are the experimental estimation of the kinetic barrier without the contribution of $\log k_0$ using folding (unfolding) rupture forces.
\textbf{(B)} $\log k_0$ is obtained overlapping the experimental data to the theoretical curve. \textit{Insets}: The value of $\log k_0$ can be underestimated (top) or overestimated (bottom) leading to a worse match between theory and experiments.
}\label{fig: analisi2}
\end{figure}

In order to estimate the attempt rate at zero force we need a theoretical model for the kinetic barrier. In the case of this work we use the Kramers theory (see section \ref{sec: kramers}). Once we have evaluated $B_{\rm eff}(f)-\log k_0$, $\log k_0$ is obtained by overlapping the theoretical model to the experimental results, as shown in Fig. \ref{fig: analisi2}. The sensitivity in the determination of $\log k_0$ is similar to the sensitivity in determining $\Delta G_N(0)$: at $\log k_0$=10.5 we find the best match, and for values 0.4 greater or smaller the result significantly worsens (insets in Fig. \ref{fig: analisi2}B).

\subsection{Sensitivity of the method at determining $\mathbf{P}$}

\begin{figure}[ht]
\centering
\includegraphics[scale=1]{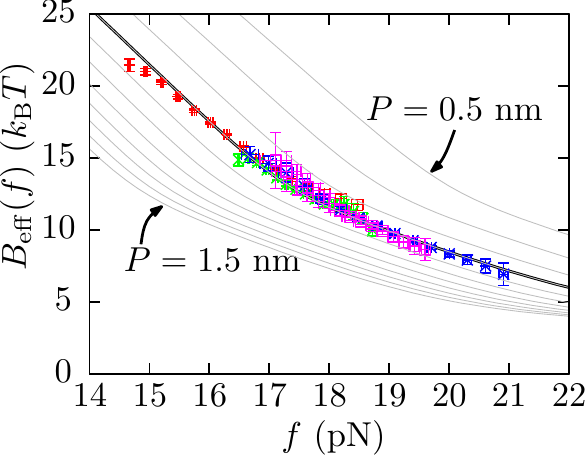}
\caption{\textbf{Determination of $\mathbf{P}$.} Data obtained at 1 mM [Mg$^{2+}$].
Gray lines are evaluations of the kinetic barrier using Kramers theory, eq. \eqref{eq: theor_effBarrier2}, for different values of the persistence length ($P=0.5,0.6,0.7,\dots,1.5$ nm). Experimental data is analyzed using also different values of the persistence length. In this case, at $P=0.8\pm0.1$ nm we obtain the best overlapping between theory and experiments.
}\label{fig: analisi3}
\end{figure}

The elastic contribution to $B_{\rm eff}(f)$ only applies to folding data, eq. \eqref{eq: BF}, and regulates the slope of the experimental estimation of the kinetic barrier in the range of experimentally measured folding forces.
In this work we determine the persistence length $P$ of ssRNA by comparing the profile of the kinetic barrier estimated from experimental data with the one evaluated using the Kramers theory (see Fig. \ref{fig: analisi3}). 
The sensitivity at determining $P$ is limited by the determination of $\Delta G_N(0)/k_{\rm B}T$, $\log k_0$ and the error bars estimated for $B_{\rm eff}^{(U/F)}(f)$. Therefore, the is a feedback in the determination of the $\Delta G_N(0)/k_{\rm B}T$, $\log k_0$ and $P$: optimal values are those that give a best fit between experimental and theoretical estimations of the kinetic barrier.

The same methodology was used to determine the value of $m$ in the non-specific correction for the free energy of formation of one base pair, where the elastic parameters are known but the free energy is unknown (see section 3.5 and Fig. 4 in the main paper): we look for the best matching between theory and experiments.

\clearpage

\section{UV Absorbance experiments}

\begin{figure}[ht]
 \centering
\includegraphics[scale=1]{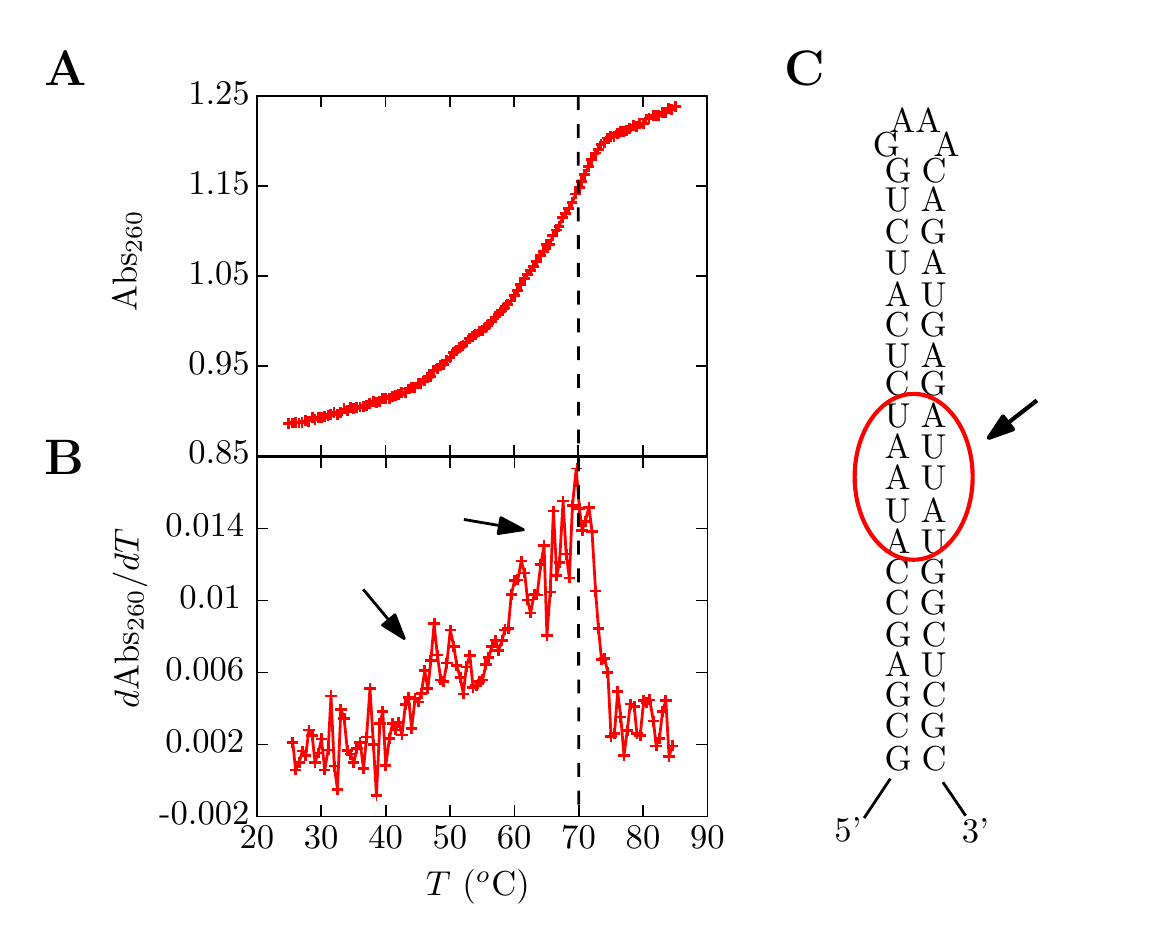}
\caption{UV Absorbance of the RNA hairpin. 
\textbf{(A)} Experimental results of the absorbance of our RNA hairpin as a function of temperature. 
\textbf{(B)} First derivative of the absorbance as a function of temperature. Its maximum (black dashed line) defines the melting temperature. 
\textbf{(C)} The RNA hairpin under study has a region with a high A-U content in the stem which could lead to premelted states.}\label{fig: melting}
\end{figure}

In order to determine the effect of salt on the stability of the hairpin we obtained the melting profile of the molecule using UV absorbance at 260 nm. The melting temperature was measured at 70$\pm1^{\rm o}$C using a buffer containing 100 mM Tris.HCl, 1 mM EDTA and no NaCl neither MgCl$_2$. Results can be observed in Fig. \ref{fig: melting}A. We calculated the first derivative of the absorbance as a function of temperature (Fig. \ref{fig: melting}B) and observed several maximums along the resulting profile (see arrows), which denote the presence of pre-melted states. For instance, regions with a richer A-U content in the middle of the stem may dissociate before the whole hairpin is unfolded (Fig. \ref{fig: melting}C). This result invalidates the two-states assumption used to extract thermodynamic parameters from the melting curve. 

We tried to obtain the melting profile adding 100 mM NaCl to the buffer, and the melting temperature was too high and the sample started boiling and evaporating before any relevant signal could be obtained. Therefore, for this RNA hairpin melting curves cannot be measured at the experimental conditions used in our pulling experiments with optical tweezers, and our results can only be compared with Mfold \cite{Walter1994,Mathews1999,Zuker2003} and other theoretical predictions \cite{Manning1978,TanChen2005,TanChen2006}

\clearpage

\section{Rupture force histograms}

\begin{figure}[ht]
\centering
\includegraphics[scale=1]{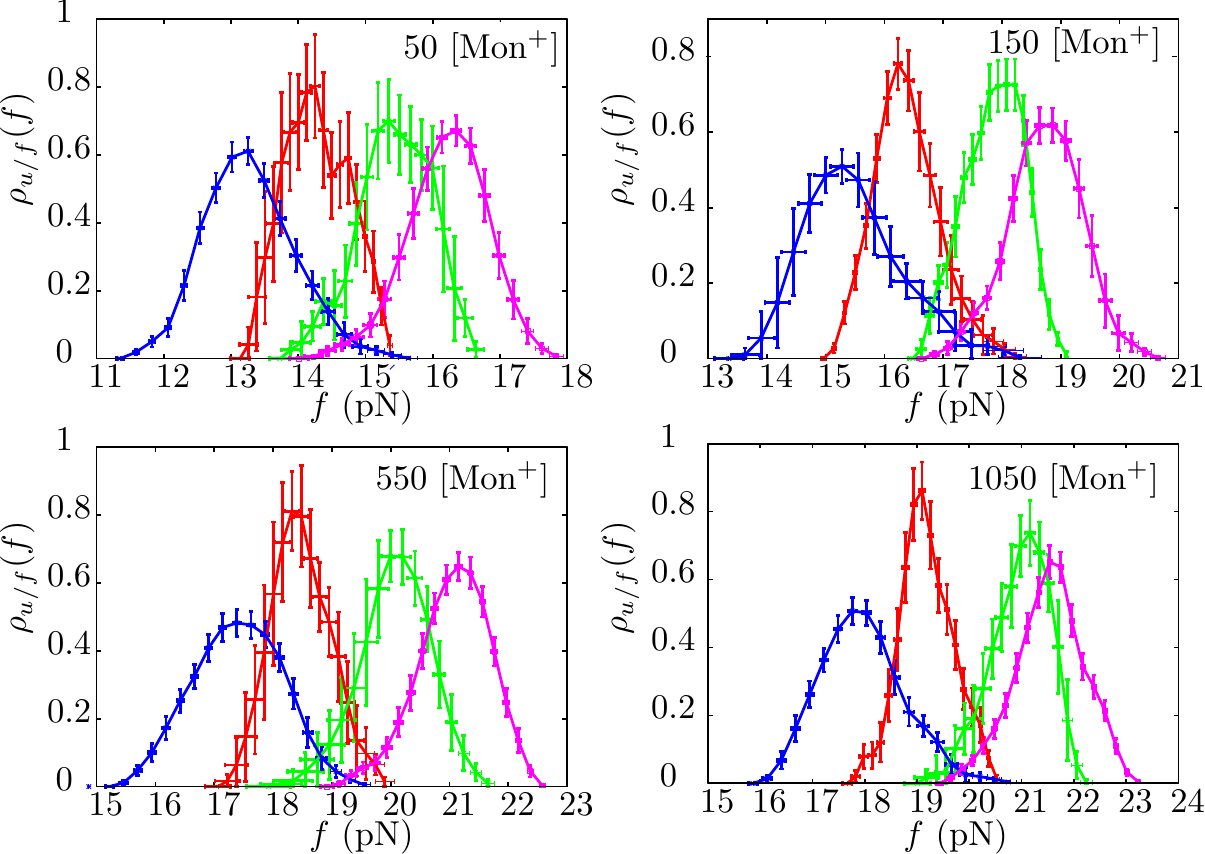}
\caption{Probability distributions of the unfolding and folding first rupture forces measured at different pulling speeds and different monovalent ionic condition. Red points are folding forces at 1.8 pN/s, green are unfolding forces at 1.8 pN/s, blue are folding forces at 12.5 pN/s and magenta are unfolding forces at 12.5 pN/s.}
\end{figure}

\begin{figure}[ht]
\centering
\includegraphics[scale=1]{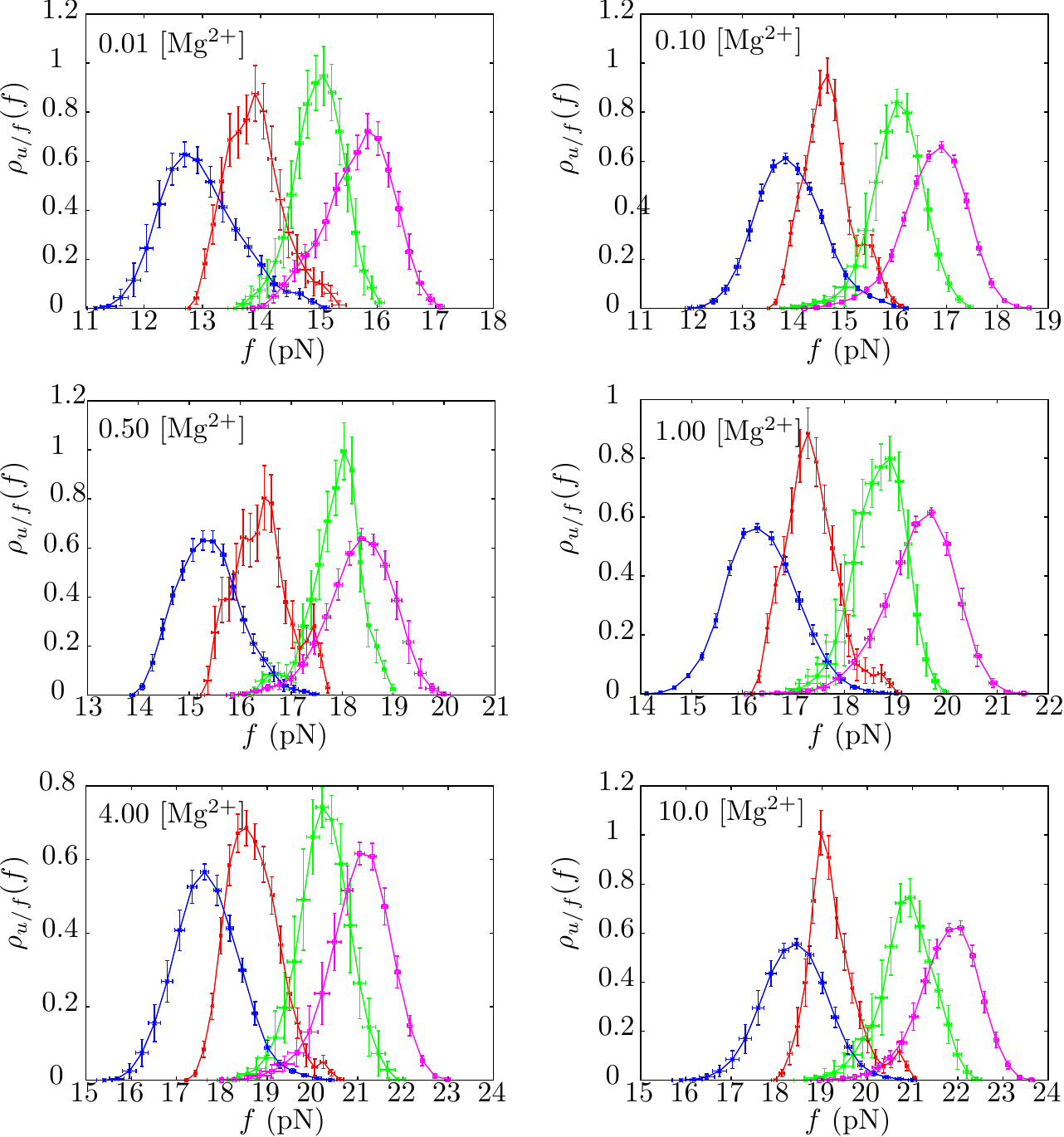}
\caption{Probability distributions of the first rupture forces measured at different pulling speeds and different magnesium concentration. Red points are folding forces at 1.8 pN/s, green are unfolding forces at 1.8 pN/s, blue are folding forces at 12.5 pN/s and magenta are unfolding forces at 12.5 pN/s.}
\end{figure}

\clearpage
              
\section{Tightly Bound Ion Model}

In this section the empirical equations of the Tightly Bound Ion (TBI, \cite{TanChen2005,TanChen2006}) model used to predict the hairpin free energies at different ionic conditions are summarized \cite{Tan2007,Tan2008}.

The external parameters, obtained from the Mfold server \cite{Walter1994,Mathews1999,Zuker2003}, are:

\begin{table}[ht]
\centering
\begin{tabular}{lr@{=}l}
\hline
Bases in the loop & $N_l$ & 4\\
Bases in the helix & $N$ & 20 \\
Hairpin diameter & $d$ & 1.7 nm\\
Interphosphate distance & $a$ & 0.6 nm\\
Enthalpy at 1 M [Mon$^+$], 0 M [Mg$^{2+}$] & $\Delta H_0$ & 199 kcal/mol \\
Entropy at  1 M [Mon$^+$], 0 M [Mg$^{2+}$] & $\Delta S_0$ & 527.16 mkcal/Kmol \\
\hline
\end{tabular}
\end{table}
\vspace{0.2cm}

In what follows $x$ is the concentration of monovalent salt and $y$ is the concentration of magnesium ions. Both parameters are given in units of M. Temperature $T$ is given in Celsius. 

The empirical set of equations are:

\begin{align}
\ln Z^l_{Mon}(x)=a_1^l(x)\log(N_l-a/d+1)+b_1^l(x) (N_l-a/d+1)^2-b_1^l(x)\\
\ln Z^{c}_{Mon}(x)=c_1^l(x) N_l-d_1^l(x)\\
a_1^l(x)=(0.02 N_l-0.026)\log(x)+0.54 N_l+0.78\\
b_1^l(x)=\left(-\frac{0.01}{(N_l+1)}+0.006\right) \log(x)-\frac{7}{(N_l+1)^2}-0.01\\
c_1^l(x)=0.07 \log(x)+1.8\\
d_1^l(x)=0.21 \log(x)+1.5\\
G^l_{Mon}(x)=-(\ln Z^l_{Mon}(x)-\ln Z^{c}_{Mon}(x))
\end{align}
\begin{align}
G^h_{Mon}(x,T)=H_0-(T+273.15) S_{Mon}(x) 0.001 \\
S_{Mon}(x)=S_0-3.22 (N-1) g_1(x) \\
g_1(x)=a_1^h(x)+b_1^h(x)/N \\
a_1^h(x)=-0.075 \log(x)+0.012 \log^2(x) \\
b_1^h(x)=0.018 \log^2(x) 
\end{align}
%
\begin{align}
G_{Mon}(x,T)=G^h_{Mon}(x,T)+G^l_{Mon}(x)
\end{align}
%
\begin{align}
\ln Z^{l}_{Mg}(y)=a_2^l(y) \log(N_l-a/d+1)+b_2^l(y) (N_l-a/d+1)^2-b_2^l(y)\\
\ln Z^{c}_{Mg}(y)=c_2^l(y) N_l-d_2^l(y) \\
a_2^l(y)=\left(-\frac{1}{N_l+1}+0.32\right) \log(y)+0.7 N_l+0.43 \\
b_2^l(y)=0.0002 (N_l+1) \log(y)-5.9/(N_l+1)^2-0.003\\
c_2^l(y)=0.067 \log(y)+2.2\\
d_2^l(y)=0.163 \log(y)+2.53\\
G^l_{Mg}(y)=-(\ln Z^{l}_{Mg}(y)-\ln Z^{c}_{Mg}(y))\\
\end{align}
%
\begin{align}
G^h_{Mg}(y,T)=H_0-(T+273.15) S_{Mg}(y) 0.001 \\
S_{Mg}(y)=S_0-3.22 (N-1) g_2(y) \\
g_2(y)=a_2^h(y)+b_2^h(y)/N^2 \\
a_2^h(y)=-0.6/N+0.025 \log(y)+0.0068 \log^2(y)\\
b_2^h(y)=\log(y)+0.38 \log^2(y)
\end{align}
%
\begin{align}
G_{Mg}(y,T)=G^h_{Mg}(y,T)+G^l_{Mg}(y)
\end{align}
%
\begin{align}
x_1^l(x,y)=x/(x+(7.2-20/N_l) (40-\log(x)) y)\\
G^l_{Mon,Mg}(x,y)=x_1^l(x,y) G^l_{Mon}(x)+(1-x_1^l(x,y)) G^l_{Mg}(y)
\end{align}
%
\begin{align}
G^h_{Mon,Mg}(x,y,T)=H_0-(T+273.15) S_{Mon,Mg}(x,y) 0.001 \\
S_{Mon,Mg}(x,y)=S_0-3.22 \left[(N-1) \left(x_1^h(x,y) g_1(x)+(1-x_1^h(x,y)) g_2(y)\right)+g_{1,2}(x,y)\right] \\
x_1^h(x,y)=\frac{x}{x+(8.1-32.4/N) (5.2-\log(x)) y} \\
g_{1,2}(x,y)=-0.6 x_1^h(x,y) (1-x_1^h(x,y)) \log(x) \log((1/x_1^h(x,y)-1) x)/N
\end{align}
%
\begin{align}
\boxed{G_{Mon,Mg}(x,y,T)=G^h_{Mon,Mg}(x,y,T)+G^l_{Mon,Mg}(x,y)}
\end{align}

Where $G_{Mon,Mg}(x,y,T)$ is the free energy at any temperature and at any monovalent and magnesium ion concentration.

\clearpage

\section{Comparison to the counterion condensation theory}

There are two successful theories to account for the energetic interactions between ions in solution and nucleic acids: the Poisson-Boltzmann theory and the counterion condensation theory derived by Manning \cite{Manning1978, Manning2002}. These theories are based on different mean field approaches and neglect any kind of correlations between the ions in the solution. More recently a new theory known as the Tightly Bound Ion (TBI) model has been introduced \cite{TanChen2005}, which accounts for the different modes of correlations between counterions. 

\begin{figure}[ht]
\centering
\includegraphics[scale=1]{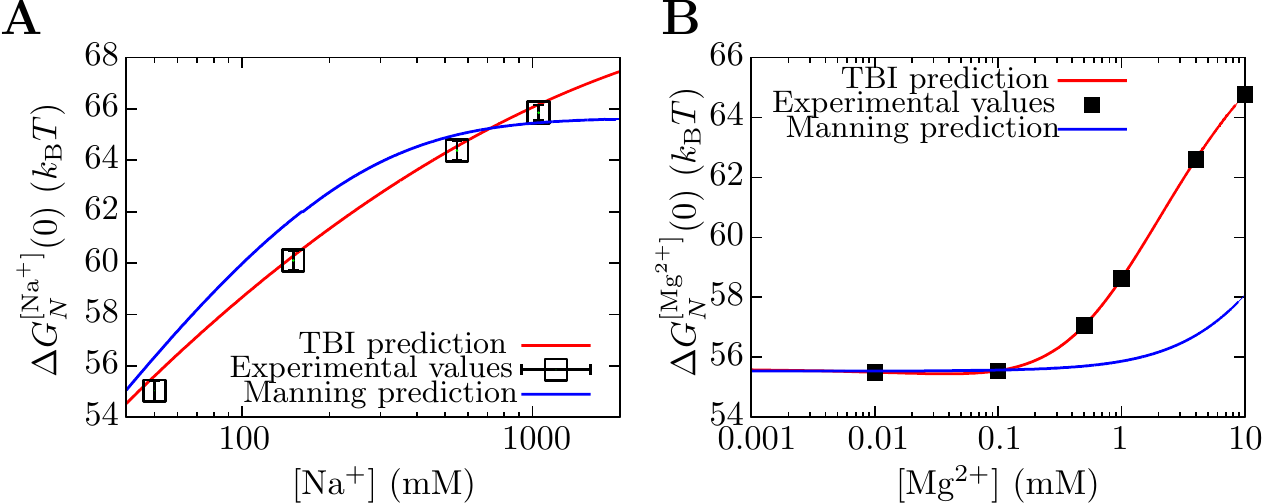}
\caption{Comparison between experimental data, the counterion condensation theory and the TBI model to predict the behavior of the free energy of formation of our RNA hairpin as a function of salt concentration.} \label{fig: Manning}
\end{figure}

In Fig. \ref{fig: Manning} we see the prediction provided by the Manning theory and the TBI model to the free energy of formation of our RNA hairpin as a function of the salt concentration. Because correlations between monovalent ions are negligible, we see that both the Manning theory and the TBI model give similar results under this condition (Fig. \ref{fig: Manning}A). However, correlations between Mg$^{2+}$ are important and the TBI model gives an improved prediction in this case (Fig. \ref{fig: Manning}B) \cite{TanChen2006}.

\begin{thebibliography}{99}

\bibitem{Berkhout1989} Berkhout B., Silverman R.H. and Jeang K.T. (1989)
Tat trans-activates the human immunodeficiency virus through a nascent RNA target. 
\textit{Cell} \textbf{59(2),} 273-282.
\bibitem{Olsthoorn1999} Olsthoorn R.C.L., Mertens S., Brederode F.T. and Bol J.F. (1999)
A conformational switch at the 3' end of a plant virus RNA regulates viral replication. 
\textit{EMBO J.} \textbf{18(17),} 4856-4864. 
\bibitem{Gollnick2005} Gollnick P., Babitzke P., Antson A. and Yanofsky C. (2005) 
Complexity in Regulation of Tryptophan Biosynthesis in Bacillus subtilis.
\textit{Annu. Rev. Genet.} \textbf{39,} 47-68.
\bibitem{Garst2009} Garst A.D. and Batey R.T. (2009)
A switch in time: Detailing the life of a riboswitch. 
\textit{Biochim. Biophys. Acta } \textbf{1789(9-10),} 584-591.
\bibitem{Li2011} Li L. and Liu Y. (2011)
Diverse Small Non-coding RNAs in RNA Interference Pathways. 
\textit{Methods in Molecular Biology} \textbf{764,} 169-182.

\bibitem{Kawaji2008} Kawaji,H. and Hayashizaki,Y. (2008) Exploration
of small RNAs. \textit{PLoS Genet.} \textbf{4(1),} e22. 
\bibitem{Zhang2010} Zhang,J., Lau,M. and Ferr\'e-D'Amar\'e,A. (2010)
Ribozymes and riboswitches: modulation of RNA function by small molecules.
\textit{Biochemistry} \textbf{49,} 9123-9131.

\bibitem{Bardaro2009} Bardaro Jr, M.F., Shajani, Z., Patora-Komisarska, K., Robinson, J.A. and Varani, G.
(2009) How binding of small molecule and peptide ligands to HIV-1 TAR alters the RNA motional landscape. 
\textit{Nucl. Acids Res.} \textbf{37,} 1529-1540.

\bibitem{Tinoco1999} Tinoco Jr I. and Bustamante C. (1999)
How RNA Folds.
\textit{J. Mol. Biol.} \textbf{293,} 271-281.
\bibitem{Zhao2010} Zhao, P., Zhang, W.-B. and Chen, S.-J. (2010) Predicting Secondary Structural Folding Kinetics
for Nucleic Acids. \textit{Biophys. J.} \textbf{98,} 1617-1625.

\bibitem{Chen2000} Chen S.J. and Dill K.A. (2000) RNA folding energy landscapes.
\textit{Proc. Natl. Acad. Sci. U. S. A.} \textbf{97(2),} 646-51. 
\bibitem{Hyeon2006} Hyeon C. and Thirumalai D. (2006) Forced-unfolding
and force-quench refolding of RNA hairpins. \textit{Biophys J.} \textbf{90(10),} 3410-27.
\bibitem{Petesheim1983} Petesheim,M. and Turner,D.H. (1983) Base-stacking
and base-pairing contributions to helix stability: thermodynamics
of double-helix formation with CCGG, CCGGp, CCGGAp, ACCGGp, CCGGUp,
and ACCGGUp. \textit{Biochemistry} \textbf{22,} 256-263.
\bibitem{Freier1986} Freier,S.M., Kierzek,R., Jaeger,J.A., Sugimoto,N.,
Caruthers,M.H., Neilson, T. and Turner,D.H. (1986) Improved free-energy
parameters for predictions of RNA duplex stability. \textit{Proc. Natl. Acad. Sci. U.S.A.} \textbf{83,} 9373-9377.

\bibitem{Walter1994} Walter,A.E., Turner,D.H., Kim,J., Lyttle,
M.H., M\"uller,P., Mathews, D.H. and
Zuker,M. (1994) Coaxial stacking of helixes enhances binding of oligoribonucleotides
and improves predictions of RNA folding. \textit{Proc. Natl. Acad. Sci. U.S.A.} \textbf{91,} 9218-9222.
\bibitem{Serra1994}Serra,M.J., Axenson,T.J. and Turner,D.H. (1994)
A model for the stabilities of RNA hairpins based on a study of the
sequence dependence of stability for hairpins of six nucleotides.
\textit{Biochemistry} \textbf{33,} 14289-14296.
\bibitem{Serra1997} Serra,M.J., Barnes,T.W., Betschart,K., Gutierrez,M.J.,
Sprouse,K.J., Riley, C.K., Stewart,L. and Temel,R.E. (1997) Improved
parameters for the prediction of RNA hairpin stability. \textit{Biochemistry} \textbf{36,} 4844-4851.
\bibitem{Xia1998} Xia,T., SantaLucia Jr.,J., Burkard,M.E., Kierzek,R.,
Schroeder,S.J., Jiao,X., Cox,C. and Turner,D.H. (1998) Thermodynamic
parameters for an expanded nearest-neighbor model for formation of
RNA duplexes with Watson-Crick base pairs. \textit{Biochemistry} \textbf{37,} 14719-14735.
\bibitem{Mathews1999} Mathews,D.H., Sabina,J., Zuker,M. and Turner,D.H.
(1999) Expanded sequence dependence of thermodynamic parameters improves
prediction of RNA secondary structure. \textit{J. Mol. Biol.} \textbf{288,} 911-940.
\bibitem{Znosko2002}Znosko,B.M., Silvestri,S.B., Volkman,H.,
Boswell,B. and Serra,M.J. (2002) Thermodynamic parameters for an expanded
nearest-neighbor model for the formation of RNA duplexes with single
nucleotide bulges. \textit{Biochemistry} \textbf{41,} 10406-10417.
\bibitem{Mathews2004}Mathews,D.H., Disney,M.D., Childs,J.L.,
Schroeder,S.J., Zuker,M. and Turner,D.H. (2004) Incorporating chemical
modification constraints into a dynamic programming algorithm for
prediction of RNA secondary structure. \textit{Proc. Natl. Acad. Sci. U.S.A.} \textbf{101,} 7287-7292.
\bibitem{Furtig2007} F\"urtig,B., Wenter,P., Reymond,L., Richter,C.,
Pitsch,S., and Schwalbe,H. (2007) Conformational dynamics of bistable
RNAs studied by time-resolved NMR spectroscopy. \textit{J. Am. Chem. Soc.} \textbf{129,} 16222-16229.

\bibitem{Furtig2010} F\"urtig B., Wenter P., Pitsch S. and Schwalbe H. (2010)
Probing mechanism and transition state of RNA refolding. \textit{ACS Chem
Biol.} \textbf{5(8),} 753-65. 

\bibitem{Williams1989}Williams,A.P., Longfellow,C.E., Freier,S.M.,
Kierzek,R. and Turner,D.H. (1989) Laser Temperature-Jump, Spectroscopic,
and Thermodynamic Study of Salt Effects on Duplex Formation by dGCATGC.
\textit{Biochemistry} \textbf{28,} 4283-4291.
\bibitem{Ma2006} Ma H., Proctor D.J., Kierzek E., Kierzek R., Bevilacqua P.C. and Gruebele M. (2006) 
Exploring the energy landscape of a small RNA hairpin.
\textit{J. Am. Chem. Soc.} \textbf{128(5),} 1523-30. 
\bibitem{Sarkar2009} Sarkar K., Meister K., Sethi A. and Gruebele M. (2009)
Fast folding of an RNA tetraloop on a rugged energy landscape detected
by a stacking-sensitive probe. \textit{Biophys. J.} \textbf{97(5),} 1418-27. 
\bibitem{Sarkar2010} Sarkar K., Nguyen D.A. and Gruebele M. (2010) Loop
and stem dynamics during RNA hairpin folding and unfolding. 
\textit{RNA} \textbf{12,} 2427-34.
\bibitem{Stancik2008} Stancik A.L. and Brauns E.B. (2008) Rearrangement
of partially ordered stacked conformations contributes to the rugged
energy landscape of a small RNA hairpin. 
\textit{Biochemistry} \textbf{47(41),} 10834-40.
\bibitem{Zhang2006} Zhang W. and Chen S.J. (2006) Exploring the complex
folding kinetics of RNA hairpins: I. General folding kinetics analysis.
\textit{Biophys. J.} \textbf{90(3),} 765-77.

\bibitem{Hyeon2005} Hyeon C. and Thirumalai D. (2005) Mechanical unfolding
of RNA hairpins. 
\textit{Proc. Natl. Acad. Sci. U. S. A.} \textbf{102(19),} 6789-94.

\bibitem{Liphardt2001} Liphardt, J., Onoa, B., Smith, S.B., Tinoco Jr., I. and Bustamante, C. (2001) Reversible 
Unfolding of Single RNA Molecules by Mechanical Force. \textit{Science} \textbf{292,} 733-737.
\bibitem{Tinoco2004} Tinoco Jr, Collin D. and Li P.T.X. (2004) The effect
of force on thermodynamics and kinetics: unfolding single RNA molecules.
\textit{Biochem. Soc. Trans.} \textbf{32(Pt 5),}  757-760.
\bibitem{Li2006} Li, P.T., Collin, D., Smith, S.B., Bustamante, C. and Tinoco, I. Jr (2006) Probing the mechanical
folding kinetics of TAR RNA by hopping, force-jump, and force-ramp methods. \textit{Biophys. J.} \textbf{90,} 250-260.

\bibitem{Bustamante2004} Bustamante C., Chemla Y.R., Forde N.R. and Izhaky
D. (2004) Mechanical processes in biochemistry. \textit{Annu. Rev. Biochem.} \textbf{73,} 705-48. 
\bibitem{Tinoco2010}Tinoco,I., Chen,G. and Qu,X. (2010) RNA reactions
one molecule at a time. \textit{Cold Spring Harb. Perspect. Biol.} \textbf{2,} a003624.

\bibitem{Manosas2006}Manosas,M., Collin,D. and Ritort,F. (2006)
Force-dependent fragility in RNA hairpins. \textit{Phys. Rev. Lett.} \textbf{96,} 218301.

\bibitem{Huguet2010}Huguet,J.M., Bizarro,C.V., Forns,N., Smith,S.B.,
Bustamante,C. and Ritort, F. (2010) Single-molecule derivation of
salt dependent base-pair free energies in DNA. \textit{Proc. Natl. Acad. Sci. U.S.A.} \textbf{107,} 15431-15436.
\bibitem{Bustamante2006}Bustamante,C. and Smith,S.B. (2006) Patent
US 7133132 B2. 

\bibitem{Record1975} Record, M. Th., Jr. (1975). Effects of Na$^{+}$ and
Mg$^{2+}$ ions on the helix-coil transition of DNA. \textit{Biopolymers} \textbf{14,} 2137-2158.
\bibitem{Record1976} Record Jr., M. T., Woodbury, C. P. and Lohman,
T. M. (1976) Na+ effects on transitions of DNA and polynucleotides
of variable linear charge density. \textit{Biopolymers} \textbf{15(5),} 893-915.
\bibitem{Williams1996} Williams,D.J. and Hall,K.B. (1996) Thermodynamic
comparison of the salt dependence of natural RNA hairpins and RNA
hairpins with non-nucleotide spacers. \textit{Biochemistry} \textbf{35,} 14665-14670.
\bibitem{Takach2004} Takach,J., Mikulecky,P. and Feig,A. (2004)
Salt-dependent heat capacity changes for RNA duplex formation. 
\textit{J. Am. Chem. Soc.} \textit{126(21),} 6530-6531.
\bibitem{Tan2005} Tan,Z.J. and Chen,S.J. (2005) 
Electrostatic correlations and fluctuations for ion binding to a finite length polyelectrolyte.
\textit{J. Chem. Phys.} \textbf{122(4),} 44903.

\bibitem{Tan2007} Tan,Z.J. and Chen,S.J. (2007) RNA helix stability
in mixed Na$^{+}$/Mg$^{2+}$ solution. \textit{Biophys. J.} \textbf{92,} 3615-3632.
\bibitem{Tan2008} Tan,Z.J. and Chen,S.J. (2008) Salt dependence
of nucleic acid hairpin stability. \textit{Biophys. J.} \textbf{95,} 738-752.

\bibitem{Heilman2001} Heilman-Miller, S. L., Thirumalai, D. and Woodson, S. A. (2001)
Role of Counterion Condensation in Folding of Tetrahymena Ribozyme. I: Equilibrium Stabilization by Cations.
\textit{J. Mol. Biol.} \textbf{306,} 1157-1166.
\bibitem{TanChen2009} Z. J. Tan and S. J. Chen (2009)
Predicting Electrostatic Forces in RNA Folding
\textit{Methods in Enzymology} \textbf{469,} Chapter 22.

\bibitem{Collin2005}Collin,D., Ritort,F., Jarzynski,C., Smith,S.B.,
Tinoco Jr,I. and Bustamante, C. (2005) Verification of the Crooks
fluctuation theorem and recovery of RNA folding free energies. \textit{Nature} \textbf{437,} 231-234.

\bibitem{McManus2002}McManus,M.T., Petersen,C.P., Haines,B.B.,
Chen,J. and Sharp,P.A. (2002) Gene silencing using micro-RNA designed
hairpins. \textit{RNA} \textbf{8,} 842-850.

\bibitem{Casey1977}Casey,J. and Davidson, N. (1977) Rates of formation
and thermal stabilities of RNA:DNA and DNA:DNA duplexes at high concentrations
of formamide. \textit{Nucleic Acids Res.} \textbf{4,} 1539-1552.

\bibitem{Owczarzy2008}Owczarzy,R., Moreira,B.G., You,Y., Behlke,M.A.
and Walder,J.A. (2008) Predicting stability of DNA duplexes in solutions
containing magnesium and monovalent cations. \textit{Biochemistry} \textbf{47,} 5336-5353.

\bibitem{Liphardt2002}Liphardt,J., Dumont,S., Smith,S.B., Tinoco,I.
and Bustamante,C. (Jun, 2002) Equilibrium information from nonequilibrium
measurements in an experimental test of Jarzynski's equality. \textit{Science} \textbf{296(5574),} 1832-1835.

\bibitem{Cocco2003}Cocco,S., Marko,J.F. and Monasson,R. (2003)
Slow nucleic acid unzipping kinetics from sequence-defined barriers.
\textit{Eur. Phys. J. E: Soft Matter Biol. Phys.} \textbf{10,} 153-161.
\bibitem{Manosas2005}Manosas,M. and Ritort,F. (2005) Thermodynamic
and kinetic aspects of RNA pulling experiments. \textit{Biophys. J.} \textbf{88,} 3224-3242.

\bibitem{Woodside2006} Woodside M.T., Anthony P.C., Behnke-Parks
W.M., Larizadeh K., Herschlag D. and Block S.M. (2006) Direct measurement of
the full, sequence-dependent folding landscape of a nucleic acid.
\textit{Science} \textbf{314(5801),} 1001-4.

\bibitem{Mossa2009}Mossa,A., Manosas,M., Forns,N., Huguet,J.M.
and Ritort,F. (2009) Dynamic force spectroscopy of DNA hairpins: I.
Force kinetics and free energy landscapes. \textit{J. Stat. Mech.} P02060.
\bibitem{Forns2011}Forns,N., de Lorenzo,S., Manosas,M., Hayashi,K.,
Huguet,J.M. and Ritort, F. (2011) Improving signal/noise resolution
in single-molecule experiments using molecular constructs with short
handles. \textit{Biophys. J.} \textbf{100(7),} 1765-1774.

\bibitem{Bustamante1994}Bustamante,C., Marko,J.F., Siggia,E.D.
and Smith,S. (1994) Entropic elasticity of lambda-phage DNA. \textit{Science} \textbf{265,} 1599-1600.
\bibitem{Marko1995}Marko,J.F. and Siggia,E.D. (1995) Stretching
DNA. \textit{Macromolecules} \textbf{28,} 8759-8770.

\bibitem{Kramers1940}Kramers,H.A. (1940) Brownian motion in a
field of force and the diffusion model of chemical reactions. \textit{Physica} \textbf{7,} 284-304.
\bibitem{Bell1978}Bell,G.I. (1978) Models for the specific adhesion
of cells to cells. \textit{Science} \textbf{200,} 618-627.
\bibitem{EvansRitchie1997}Evans,E. and Ritchie,K. (1997) Dynamic
strength of molecular adhesion bonds. \textit{Biophys. J.} \textbf{72,} 1541-1555.
\bibitem{Qian1999}Qian,H. and Shapiro,B.E. (1999) Graphical Method
for Force Analysis: Macromolecular Mechanics With Atomic Force Microscopy.
\textit{Proteins: Structure, Function, and Genetics} \textbf{37,} 576-581.
\bibitem{Zwanzig2001}Zwanzig,R. (2001) Nonequilibrium Statistical
Mechanics, Oxford University Press, 1st edition. 
\bibitem{Hyeon2007} Hyeon, C. and Thirumalai, D. (2007) Measuring the energy landscape roughness and the transition
state location of biomolecules using single molecule mechanical unfolding experiments. \textit{J. Phys.: Condens. Matter} \textbf{19,} 113101.

\bibitem{Evans2009} Evans,E., Halvorsen,K., Kinoshita,K. and Wong,W.P.
(2009) A new approach to analysis of single-molecule force measurements.
\textit{Springer Science+ Business Media}

\bibitem{Tinoco1971} Tinoco,I., Uhlenbeck,O.C. and Levine,M.D.
(1971) Estimation of Secondary Structure in Ribonucleic Acids19. \textit{Nature} \textbf{230,} 362-367.
\bibitem{Zuker2003}Zuker,M. (2003) Mfold server for nucleic acid
folding and hybridization prediction. \textit{Nucleic Acids Res.} \textbf{31,} 3406-3415.

\bibitem{Seol2004}Seol,Y., Skinner,G.M. and Visscher,K. (2004)
Elastic properties of a singlestranded charged homopolymeric ribonucleotide.
\textit{Phys. Rev. Lett.} \textbf{93,} 118102.
\bibitem{Toan2006}Toan,N.M. and Micheletti,C. (2006) Inferring
the effective thickness of polyelectrolytes from stretching measurements
at various ionic strengths: applications to DNA and RNA. \textit{J. Phys.: Condens. Matter.} \textbf{18,} S269-S281.

\bibitem{ChenPollack2012} Chen, H., Meisburger, S. P., Pabit, S. A., Sutton, J. L., Webb, W. W. and Pollack, L. (2012)
Ionic strength-dependent persistence lengths of single-stranded RNA and DNA.
\textit{Proc. Natl. Acad. Sci. U.S.A.} \textbf{109,} 799-804.

\bibitem{Micka1996}Micka,U. and Kremer,K. (1996) Persistence
length of the Debye-H\"uckel model of
weakly charged flexible polyelectrolyte chains. \textit{Phys. Rev. E: Stat., Nonlinear, Soft Matter Phys.} \textbf{54,} 2653-2662.

\bibitem{SantaLucia1998}SantaLucia,J.J. (1998) A unified view
of polymer, dumbbell, and oligonucleotide DNA nearest-neighbor thermodynamics.
\textit{Proc. Natl. Acad. Sci. U.S.A.} \textbf{95,} 1460-1465.
\bibitem{SantaLucia2004}SantaLucia,J.J. and Hicks,D. (2004) The
Thermodynamics of DNA structural motifs. \textit{Annu. Rev. Biophys. Biomol. Struct.} \textbf{33,} 415-440.
\bibitem{Shkel2004}Shkel,I.A. and Record,J.M.T. (2004) Effect
of the number of nucleic acid oligomer charges on the salt dependence
of stability ($\Delta G_{37^{o}}$) and melting temperature ($T_{m}$):
NLPB analysis of experimental data. \textit{Biochemistry} \textbf{43,} 7090-7101.

\bibitem{Vieregg2007}Vieregg,J., Cheng,W., Bustamante,C. and
Tinoco,I.J. (2007) Measurement of the effect of monovalent cations
on RNA hairpin stability. \textit{J. Am. Chem. Soc.} \textbf{129,} 14966-14973.

\bibitem{Peyret2000}Peyret,N. (2000) Prediction of Nucleic Acid Hybridization:
Parameters and Algorithms. \textit{PhD thesis}. Wayne State University Department
of Chemistry, Detroit, MI.

\bibitem{Dove1962} Dove, W. F. and Davidson, N. (1962) Cation effects
on the denaturation of DNA. \textit{Journal of Molecular Biology} \textbf{5(5),} 467-478.
\bibitem{Privalov1969} Privalov, P. L., O. B. Ptitsyn, and T.
M. Birshtein. (1969) Determination of stability of the DNA double helix
in an aqueous medium. \textit{Biopolymers} \textbf{8,} 559-571.

\bibitem{Manning1972} Manning, G. S. 1972. On the application
of polyelectrolyte ``limiting laws''
to the helix-coil transition of DNA. I. Excess univalent cations.
\textit{Biopolymers} \textbf{11,} 937-949.
\bibitem{Manning1978} Manning, G. S. (1978)
The molecular theory of polyelectrolyte solutions with applications to the electrostatic properties of polynucleotides.
\textit{Quart. Rev. biophys.} \textbf{11,} 179-246.
\bibitem{Record1978} Record, M. T., Jr, Anderson, C. F. and Lohmna, T. M. (1978)
Thermodynamic analysis of ion effects on the binding and conformational equilibra of proteins and nucleic acids: the roles of ions association of release, secreening, and ion effects on water activity.
\textit{Quart. Rev. Biophys.} \textbf{11,} 103-178.

\bibitem{Hammond1955} Hammond G.S. (1955) A correlation of reaction rates.
\textit{J. Am. Chem. Soc.} \textbf{77,} 334-338.

\bibitem{Owczarzy2004}Owczarzy,R., You,Y., Moreira,B.G., Manthey,J.A.,
Huang,L., Behlke,M.A. and Walder,J.A. (2004) Effects of sodium ions
on DNA duplex oligomers: improved predictions of melting temperatures.
\textit{Biochemistry} \textbf{43,} 3537-3554.
\bibitem{Jost2009}Jost,D. and Everaers,R. (2009) A unified Poland-Scheraga
model of oligoand polynucleotide DNA melting: salt effects and predictive
power. \textit{Biophys. J.} \textbf{96,} 1056-1067.

\bibitem{TanChen2006} Tan, Z. J. and Chen, S. J. (2006)
Nucleic Acid Helix Stability: Effects of Salt Concentration, Cation Valence and Size, and Chain Length.
\textit{Biophys. J.} \textbf{90,} 1175-1190.

\bibitem{Manning2002} Manning, G. S. (2002)
Electrostatic free energy of the DNA double helix in counterion condensation theory.
\textit{Biphys. Chem.} \textbf{101-102,} 461-473.

\bibitem{Rich2003}Rich,A. (2003) The double helix: a tale of
two puckers. \textit{Nat. Struct. Mol. Biol.} \textbf{10,} 247-249.

\bibitem{Cole1972}Cole,P.E., Yang,S.K. and Crothers,D.M. (1972)
Conformational changes of transfer ribonucleic acid. Equilibrium phase
diagrams. \textit{Biochemistry} \textbf{11,} 4358-4368.

\bibitem{Schroeder2000}Schroeder,S.J. and Turner,D.H. (2000)
Factors Affecting the Thermodynamic Stability of Small Asymmetric
Internal Loops in RNA. \textit{Biochemistry} \textbf{39,} 9257-9274.
\bibitem{Shankar2006}Shankar,N., Kennedy,S.D., Chen,G., Krugh,T.R.
and Turner,D. H. (2006) The NMR Structure of an Internal Loop from
23S Ribosomal RNA Differs from Its Structure in Crystals of 50S Ribosomal
Subunits. \textit{Biochemistry} \textbf{45,} 11776-11789.

\bibitem{Shklovskii1999} Shklovskii, B. I. (1999)
Screening of a macroion by multivalent ions: Correlation-induced inversion of charge.
\textit{Phys. Rev. E} \textbf{60 (5),} 5082-5811.

\bibitem{Douglas2009} Douglas, S.M., Dietz,H., Liedl,T., H\"ogberg,B.,
Graf,F. and Shih,W.M. (2009) Self-assembly of DNA into nanoscale three-dimensional
shapes. \textit{Nature} \textbf{459,} 414-418.

\bibitem{Cain1997} Cain R. J. and Glick G. D. (1997) The effect of cross-links on the conformational
dynamics of duplex DNA. Nucl. Acids Res. 25(4):836-842.

\bibitem{Arrhenius} Arrhenius, S. (1889)
On the reaction rate of the inversion of non-refined sugar upon souring.
\textit{Z Phys Chem} \textbf{4,} 226–248. 

\bibitem{TanChen2005} Tan, Z. J. and Chen, S. J. (2005)
Electrostatic correlations and fluctuations for ion binding to a finite length polyelectrolyte.
\textit{J. Chem. Phys.} \textbf{122,} 044903.

\end{thebibliography}
\end{document}